\documentclass[
  journal=pasa,
  manuscript=article-type,
  year=2020,
  volume=37,
]{cup-journal}
\usepackage{hyperref}
\hypersetup{
    colorlinks = true,
    allcolors  = {blue} 
}
\usepackage{amsmath}
\usepackage[nopatch]{microtype}
\usepackage{booktabs}
\usepackage{appendix}
\usepackage{amssymb}

\usepackage{caption}

\usepackage[T1]{fontenc}
\usepackage[normalem]{ulem}
\usepackage{xcolor}

\newcommand{\ms}{M$_{\odot}$}
\newcommand{\rs}{R$_{\odot}$}

\defcitealias{Nibbs2025}{Paper~I}

\title{Weakened Inspirals --- II: High Mass Ratio Common Envelope Interactions in AGB Stars}

\author{J. Nibbs}
\email[J. Nibbs]{jack.nibbs@hdr.mq.edu.au}
\affiliation{School of Mathematical \& Physical Sciences, Macquarie University, Macquarie Park, Sydney 2109, NSW, Australia}
\alsoaffiliation{Astrophysics and Space Technologies Research Centre, Macquarie University, Sydney, 2109 NSW, Australia}

\author{Orsola De Marco}
\affiliation{School of Mathematical \& Physical Sciences, Macquarie University, Macquarie Park, Sydney 2109, NSW, Australia}
\alsoaffiliation{Astrophysics and Space Technologies Research Centre, Macquarie University, Sydney, 2109 NSW, Australia}

\author{Lionel Siess}
\affiliation{Institut d’Astronomie et d’Astrophysique, Universit\'{e} Libre de Bruxelles (ULB, BLU), CP 226, B-1050 Brussels, Belgium}

\author{Miguel Gonzalez-Bolivar}
\affiliation{Astrophysics and Space Technologies Research Centre, Macquarie University, Sydney, 2109 NSW, Australia}
\alsoaffiliation{Australian Astronomical Optics, Macquarie University, 9 Wally’s Walk, Macquarie University,
NSW 2109, Australia}

\author{Ryosuke Hirai}
\affiliation{RIKEN Cluster for Pioneering Research (CPR), RIKEN, Wako, Saitama 351-0198, Japan}
\alsoaffiliation{School of Physics and Astronomy, Monash University, VIC 3800, Australia}
\alsoaffiliation{OzGrav: The ARC Centre of Excellence for Gravitational Wave Discovery, Australia}

\author{Daniel J. Price}
\affiliation{School of Physics and Astronomy, Monash University, VIC 3800, Australia}
\alsoaffiliation{IPAG, Univ. Grenoble Alpes, CNRS, 38000 Grenoble, France}


\begin{document}

\begin{abstract}
Post-red and asymptotic giant branch (RGB and AGB, respectively) binaries exhibit orbital periods and eccentricities that indicate that they must have undergone a strong binary interaction when they were on the RGB or AGB. Yet the distribution of their orbital elements challenge mass transfer and interaction theory. In particular, many systems must have experienced a strong binary interaction involving mass transfer and orbital period reduction, without undergoing a deep inspiral. In this work, we present three-dimensional smoothed particle hydrodynamics simulations of CE interactions involving an AGB star and a range of main sequence companions, extending our previous study that concentrated on RGB binaries. We explore four mass ratios, $q\equiv M_2/M_1 = 0.35,0.78,1.00,1.18, 1.50$, and two equations of state: ideal gas and a tabulated equation of state that includes recombination energy. We find that AGB donors produce substantially larger post-interaction periods than equivalent RGB systems, reaching $181$~\rs\ ($P=176$~days) for $q=1.18$ (tabulated equation of state), reproducing the lower range of the separations of observed post-AGB binaries. For the same mass ratio, however, ideal gas simulations no longer undergo inspiral in the simulated time of $\sim 35$~yr, denoting a far greater degree of mass transfer stability.  The relatively little mass lost via the $L_2$ and $L_3$ Lagrange points would form a very low mass circumbinary disk, but this would be overwhelmed by ejecta promoted by the inspiral. For binaries that go though a substantial inspiral, a circumbinary disk can form by fall-back of bound ejecta. The mass of this returning material is not well constrained but is approximately $0.01-0.1$~\ms. The fall-back timescales are up to a few hundred years. The fallback material circularises around radii of $\sim 0.5$--5\ au. Our results also suggest that the envelope mass, and the detailed donor structure plays a more significant role than mass ratio alone in determining the stability and outcome of giant-star CE interactions.

\end{abstract}

\section{Introduction}\label{sec:Intro}

\begin{figure}
    \centering
    \includegraphics[width=\linewidth]{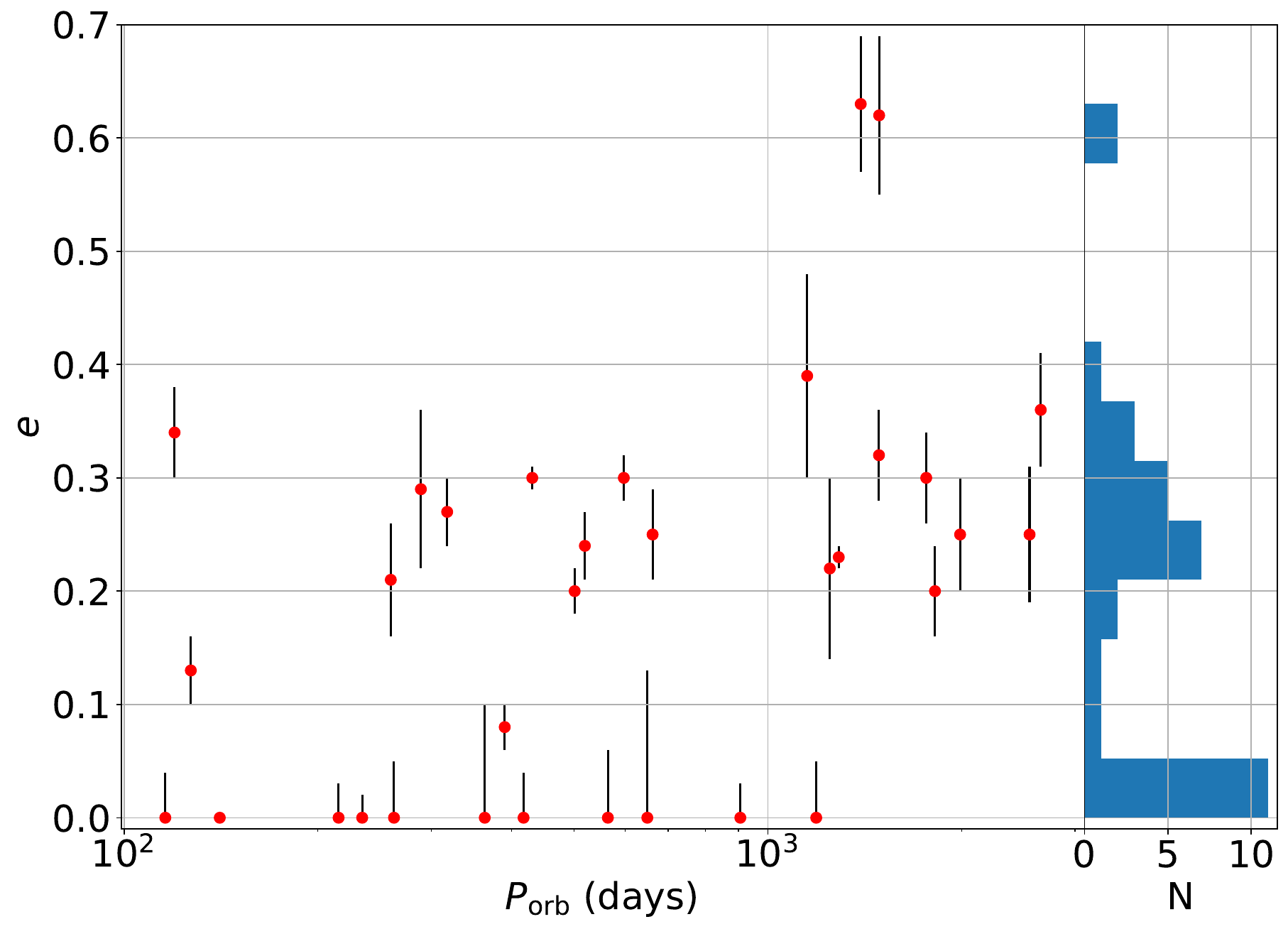}
    \caption{Post-RGB/AGB binary, observationally-derived period-eccentricity distribution. Data plotted in red was taken from \citet{kluska_population_2022} with uncertainties plotted in black from \citet{oomen_orbital_2018}. We also plot a histogram for the observation counts (right panel), which exhibits a bimodal distribution with peaks near $e\approx 0$, and $e\approx0.2-0.4$.
    }
    \label{fig:PvsE_Obs}
\end{figure}

The common envelope (CE) interaction is a phase of binary evolution crucial to explaining observed phenomena, such a short period, evolved binaries, cataclysmic variables, X-ray binaries, and a host of other objects \citep{Paczynski1971,Ivanova2013,DeMarco2017}. In a standard CE interaction, unstable Roche lobe overflow (RLOF) leads to the rapid engulfment of the companion, a dynamical inspiral, and the possible ejection of the donor’s envelope, producing a close, circularised binary or a merger. While this framework explains many short period, post-CE systems, it struggles to account for a growing population of post-RGB and post-AGB binaries with periods of 100–2000 days, and eccentricities $e\lesssim0.6$ \citep[Figure~\ref{fig:PvsE_Obs};][]{vanWinckel2009,oomen_orbital_2018,kluska_population_2022}. 

The existence of these systems poses a challenge: their semi-major axes are small enough to imply a past phase of strong binary interaction, yet wide and eccentric enough to be inconsistent with the outcome of a standard CE inspiral, which is expected to produce separations $\lesssim30$~\rs\ and nearly circular orbits \citep[e.g.,][]{Passy2012,GonzalezBolivar2022}. Moreover, essentially all known post-AGB binaries are surrounded by long-lived circumbinary discs with typical masses of $\sim10^{-2}$~\ms\ \citep{Bujarrabal2013,Hillen2017,oomen_modelling_2019}, and show photospheric depletion patterns and collimated jets indicative of  ongoing accretion onto the binary stars \citep{vanWinckel2009,Bollen2017,oomen_modelling_2019,kluska_population_2022}. These observational properties suggest a non-standard interaction history in which envelope removal, orbital evolution, and disc formation occur without a catastrophic inspiral.

Several alternative or complementary interaction channels have been proposed to explain post-RGB and post-AGB binaries, including grazing envelope evolution \citep{Soker2015}, jet-assisted envelope removal \citep{Kashi2018}, and conservative mass transfer \citep{Moltzer2025}. In particular, wind Roche-lobe overflow has been shown to enable efficient mass transfer at larger separations and, in some circumstances, to modify orbital eccentricity through phase-dependent mass loss \citep{Saladino2019,Krynski2025}. While these mechanisms may help preserve or even pump eccentricity, they do not yet fully reproduce the observed period--eccentricity distribution of post-AGB systems, especially at periods below $\sim$2000 days. 

\citet{oomen_modelling_2019} found that the distribution of companion masses for post-RGB and post-AGB binaries peaks near a mass of 1.1~\ms. This is a large value when compared to the masses of post-CE binaries, which are typically $\sim0.5$~\ms\ \citep[e.g.,][]{Iaconi2019b}. Such large companion masses may indicate that many post-AGB systems experienced interactions with mass ratios $q\equiv M_2/M_1$ close to unity, where $M_1$ is the primary's mass at the time of interaction.

\begin{figure}[h]
    \centering
    \includegraphics[width=0.65\linewidth]{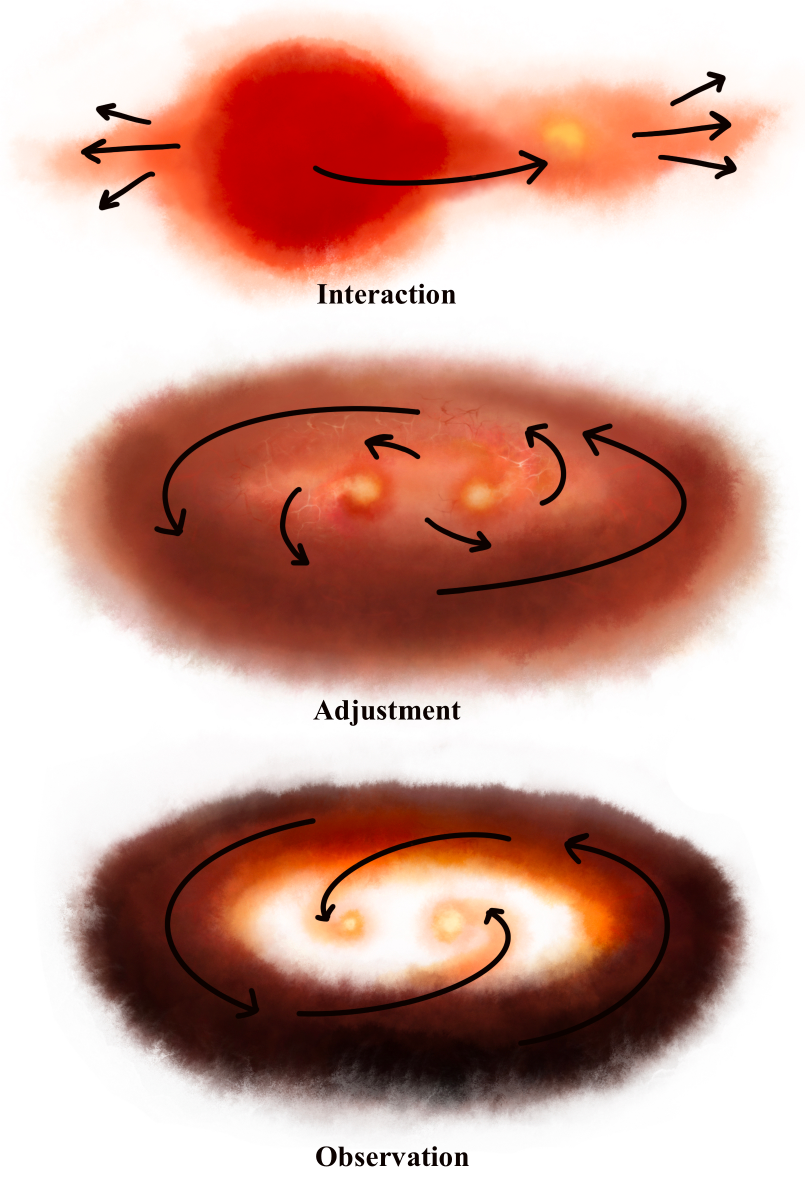}
    \caption{A schematic of the evolutionary stages leading to the formation  of post-RGB/AGB circumbinary disc systems. Stage I: an initial binary interaction removes the donor's envelope while avoiding the strong orbital shrinkage typically associated with a standard CE interaction. Stage II: following envelope ejection and orbital adjustment, the system enters a transitional phase in which the binary clears a central cavity from the surrounding bound material while a stable circumbinary disc assembles from bound circumbinary gas ejected via $L_{2/3}$ and/or fallback of ejecta. Stage III: the final long-lived configuration corresponds to the observed post-RGB/AGB binary systems, in which the circumbinary disc has largely settled into a quasi-steady structure with a cleared inner cavity. Accretion streams persist through the cavity and leave signatures of accretion onto the primary and companion.}
    \label{fig:Timeline}
\end{figure}

\citet[][hereafter, \citetalias{Nibbs2025}]{Nibbs2025} explored such high mass ratios as a way to weaken the CE interaction between a 0.88~\ms, 90~\rs\ RGB donor and a main sequence companion. Using three-dimensional smoothed particle hydrodynamics (SPH) simulations, they demonstrated that increasing the mass ratio towards and above unity increases the stability of the early mass-transfer phase, and prolongs the pre-inspiral evolution, possibly contributing to a weakening of the subsequent inspiral. Although final separations were at the small end of the observed distribution in post-RGB binaries, it suggested that at high $q$ we maybe probing the transition from a more classical CE to a more stable mass transfer where the orbital evolution is no longer dominated by gravitational friction but instead by a weaker orbital reduction mechanism such as $L_2/L_3$ mass loss and tides. The simulations in Paper I also showed that there are at least two ways of making circumbinary discs: i) $L_2$ outflows and ii) fall-back gas, with fall-back gas being more likely in the context of the simulated systems. It is indeed possible that a disc formed in this way may be left behind to become, after adjustments, the disc observed around the binaries today (see  Figure~\ref{fig:Timeline}). 

The present work extends our investigation to AGB donors. AGB stars differ fundamentally from RGB stars in both size and internal structure, possessing more extended, lower-density envelopes. These properties are expected to strongly influence the stability of mass transfer, the onset and character of any inspiral, and the efficiency with which envelope material can be ejected or lifted into circumbinary orbits \citep{Tout1991,Ge2015_II,Reichardt2019,GonzalezBolivar2022}. By modelling AGB donors that are more representative of post-AGB progenitors, we now aim to complete the picture of the extent to which inspirals weakened by a high mass ratio, can naturally produce wider post-interaction binaries and leave behind sufficient bound material to form circumbinary discs.

In this second paper of the series, we present a suite of 9 SPH simulations of CE interactions involving a thermally pulsating AGB donor with mass ratios spanning $q = 0.35 - 1.50$. Adopting the methodology of Paper I, in Section~\ref{Sec:Methods} we provide details of the simulations. In Section~\ref{sec:Results} we examine how the mass ratio influences the stability of the interaction and the degree of orbital shrinkage (Section~\ref{ssec:orbital_evolution}); the properties of the unbound and bound envelope (Section~\ref{ssec:envelope_unbinding} and \ref{ssec:bound_mass_distribution_and_properties}); the mass loss through the outer Lagrangian points (Section~\ref{ssec:pre_ce_mass_loss_through_L2_and_L3}); and, finally, the properties of the fallback gas (Section~\ref{subsec:discformation}). In Section~\ref{Sec:Summary} we summarise the results. We conclude and discuss in Section~\ref{sec:conclusion_and_discussion}.

\section{Methods}\label{Sec:Methods}

We follow a similar methodology to \citetalias{Nibbs2025}. We have performed nine simulations using the SPH (\citealt{Gingold1977,Lucy1977}) code \textsc{phantom} \citep{Price2018}. In each simulation our donor star is the same as that used by \citet{GonzalezBolivar2022} and \citet[][which included dust nucleation]{Bermudez2025}, a 2~\ms\ main sequence star evolved to the thermally-pulsating AGB till it has a mass $M_1=1.7$~\ms, and a radius of approximately $R=260$~\rs. The core of the donor is modelled as a point mass of $M_c = 0.56$~\ms, with a softened potential, $h_{\rm soft}$ (Table~\ref{tab:simulation_summary}). The envelope was subsequently relaxed following the procedure described by \citet{GonzalezBolivar2022}.

Our nine simulations include four mass ratios $q \equiv M_2/M_1 = 0.35, 0.78, 1, 1.18, 1.50$. The $q=1.18$ simulation represents a binary that initially had a mass ratio close to unity when both stars were on the main sequence. The primary was slightly more massive, allowing it to evolve more rapidly and traverse the RGB and AGB phases while the companion remained on the main sequence, consistent with observations \citep{kluska_population_2022,oomen_orbital_2018,vanWinckel2009}. During this evolution, the primary loses mass, resulting in the mass ratio adopted at the beginning of the simulation. The $q=1.50$ simulation, similarly to \citetalias{Nibbs2025}, is used as a test of stability for the interaction, as the evolutionary pathways that could lead to such a system imply the companion is a neutron star at odds with observations of post-AGB binaries.

Each simulation was performed with an ideal gas equation of state (EoS), as well as a tabulated EoS, which includes recombination energy derived from tables compiled for the \textsc{mesa} 1D code \citep{Paxton2011, Paxton2013, Paxton2015}. Recombination energy is assumed to be thermalised immediately. As such, this EoS increases the thermal energy budget of the gas and thereby enhances mass unbinding relative to an ideal gas EoS \citep[see ][]{Reichardt2020,gonzalez-bolivar_common_2022,Lau2022}. We expect that the two behaviours bracket reality in that the ideal gas EoS likely underestimates envelope unbinding, while the thermalisation of the entire recombination energy payload likely overestimates it.

All our simulations were performed with 1.3 million SPH particles, corresponding to the highest resolution considered by \citet{GonzalezBolivar2022}, who also carried out resolution tests. Our lowest mass ratio ($q=0.35$) simulations with ideal gas and tabulated EoS correspond to simulations I-hi and M-hi of  \citet{GonzalezBolivar2022}, respectively (see their table 3). 

Each one of our nine simulations starts with a circular orbit with no rotation, where the semi-major axis, $a_0$, is set such that the radius of the donor star is equal to its volume-equivalent Roche lobe radius, so mass transfer starts immediately. This radius is found using the formula from \citet{Eggleton1983},

\begin{equation}\label{eq:eggleton}
    {\frac{R_{\textrm{L}}}{a}} = \frac{0.49q^{2/3}}{0.6q^{2/3} + \textrm{ln}(1+q^{1/3})}.
\end{equation}
Our two highest $q$, ideal EoS simulations, 100I and 118I remain very stable and were stopped before any significant orbital evolution took place. A similar behaviour was also reported for the $q=1.5$, ideal EoS RGB simulation in \citetalias{Nibbs2025}. Owing to the slow evolution of these systems and the high numerical resolution adopted here, extending the simulations further would have been computationally prohibitive.

We have adopted artificial conductivity parameter values (see section 2.2.8 of \citealt{Price2018}) of unity for the ideal gas simulations, and 0.1 for tabulated EoS simulations (see Table~\ref{tab:simulation_summary}). Values larger than zero are  set to ensure the accurate energetic treatment of contact discontinuities \citep{Price2008}. The exact value of $\alpha_u$, calibrated from Kelvin-Helmoltz instabilities tests by \citet{Tricco2019} remains uncertain but \citet{GonzalezBolivar2022} found that larger values produce a modest but undesirable expansion of the star over the timescales relevant to this study. We therefore adopted a lower value, $\alpha_u=0.1$ for the tabulated EoS simulations. 

A closer comparison between RGB and AGB simulations may have included an AGB star with the same mass, or the same envelope mass as the RGB star, and identical mass ratios. However, the wallclock times of these simulations, extending to several months, made us choose simulations that were already, at least partially, in hand, relinquishing full control over the parameter sets. Additionally, observations of post-AGB binaries indicate that their main sequence progenitors' masses are closer to 2~\ms\ than 1~\ms, making the AGB mass used here closer to that of the observations than for the RGB donor of \citetalias{Nibbs2025}.

\begin{table}
    \centering
    \begin{tabular}{cccccccc}
    \hline
         Simulation&  $q$ & $M_2$&$a_0$ & $h_{\rm soft}$&$t_{\rm end}$ & $\alpha_{\rm u}$& $\alpha_{\rm min}$\\
         Name& 
     & (\ms)& (\rs)& (\rs)&(yr) & & \\
     \hline
 35I& 0.35 & 0.60 & 550 & 2.0 &27.4  & 1& 0\\
 35M& 0.35 & 0.60 & 550 & 2.0 &19.2& 0.1& 0.1\\
 78I& 0.78 & 1.33 & 650 & 2.0 &44.1  & 1& 0\\
 78M& 0.78 & 1.33 & 650 & 2.5 &15.2  & 0.1& 0.1\\
 100I& 1.00 & 1.70 & 700 & 2.0 &36.1 & 1& 0\\
 100M& 1.00 & 1.70 & 700 & 2.5 &15.2 & 0.1& 0.1\\
 118I& 1.18 & 2.00 & 725 & 2.0 &35.3 & 1& 0\\
 118M& 1.18 & 2.00 & 725 & 2.5 &15.2 & 0.1& 0.1\\
 150M& 1.50& 2.55& 765& 2.5& 21.7& 0.1&0.1\\
 \hline
 \multicolumn{6}{l}{I \& M denote ideal and tabulated (\textsc{Mesa}) EoS' respectively.} & & \\
 \end{tabular}

    \caption{Initial parameters for the 9 simulations. The mass ratio, $q \equiv M_2/M_1$, includes a donor with mass $M_1 = 1.7$ \ms, and an accretor point mass with mass $M_2$. At the initial orbital separation, $a_0$, the donor fills its Roche lobe. Other parameters are as follows: $h_{\rm soft}$ is the gravitational softening length for both the donor and accretor; $t_{\rm end}$ is the total time simulated; $\alpha_{u}$ and $\alpha_{\rm min}$ represent the artificial conductivity, and the minimum artificial viscosity, which modify \textsc{phantom}'s energy and momentum equations, respectively. Simulations are carried out with $1.372 \times 10^6$ SPH particles, with a constant particle mass, $m_{\rm p} = 8.33\times10^{-7}$~\ms.}
    \label{tab:simulation_summary}
\end{table}

\section{Results}\label{sec:Results}

In this section we determine whether increasing the mass ratio, $q$, can weaken or delay inspirals in AGB binaries in the same manner as found for the RGB binaries of \citetalias{Nibbs2025}, and whether the resulting bound ejecta naturally produces post-AGB-like circumbinary environments. We begin by presenting the orbital evolution in each simulation, and investigate the properties of the ejected material, as well as the degree to which the donor's envelope is unbound. We then look at the properties of the material ejected prior to the dynamical inspiral, and finally, we consider the possibility that the ejected, but still bound, material falls back onto the binary, and what type of circumbinary disc would form.

\begin{table*}
    \centering
    \begin{tabular}{ccccccccccccc}
\toprule
   Simulation &   $a_0$&    $t_{\rm i}$&   $a_{\rm i}$&    $t_{\rm f}$&   $a_{\rm f}$ &$P_{\rm f}$&$\langle e_{\rm f-end}\rangle$&  $t_{\rm steep}$&  $a_{\rm steep}$&  $\tau_{\rm steep}$ &$\langle \dot{a}_{\rm f-end}\rangle$&   $\dot{M}_{\rm {L_1}, 1yr}$\\
 Name & (\rs) & (yr) & (\rs) & (yr) & (\rs)  &(days)&& (yr) & (\rs) & (yr)  &(\rs~yr$^{-1}$)&(\ms~yr$^{-1}$) \\
\midrule
   35I &  550 &  14.5 &  428 &  17.7 &   28  &16 &$2.3\times 10^{-2}$&    15.3 &     152 &       0.3  &$-0.7$&  $4.6\times 10^{-4}$\\
   35M &  550 &   1.6 &  523 &   7.1 &   40  &27&$3.9\times 10^{-3}$&     4.6 &     250 &       1.4  &$0.7$&  $7.8\times 10^{-3}$\\
   78I &  650 &  24.1 &  410 &  27.8 &   81  &61&$3.9\times 10^{-2}$&    25.3 &     169 &       0.4  &$-1.0$&  $9.5\times 10^{-4}$\\
   78M &  650 &   2.4 &  615 &   7.6&  102 &86&$1.7\times 10^{-2}$&     5.7&     236&       0.9  &$-2.1$&  $4.6\times 10^{-3}$\\
  100I &  700 &   $-$ &  $-$ &   $-$&  645\textdagger &1262\textdagger&$3.3\times 10^{-3}$&    $-$&     $-$&      $-$ &$-1.5$&  $8.2\times 10^{-4}$\\
  100M &  700 &   2.5 &  669 &   8.8 &  141  &129&$1.8\times 10^{-3}$&     6.9 &     275 &       1.1  &$-2.7$&  $4.6\times 10^{-3}$\\
  118I &  725 &   $-$ &  $-$ &   $-$&  680\textdagger &1284\textdagger&$3.5\times 10^{-3}$&    $-$&     $-$&      $-$ &$-1.3$&  $9.1\times 10^{-4}$\\
  118M &  725 &   2.2 &  706 &   9.5 &  181  &176&$1.2\times 10^{-2}$&     7.2 &     344 &       1.7  &$-3.9$&  $5.0\times 10^{-3}$\\
 150M  & 765  &    2.1& 752  & 14.0& 210 &200&$6.9\times 10^{-3}$& 7.7& 453& 3.8& $-4.0$&$6.0\times 10^{-3}$\\
\bottomrule
\multicolumn{13}{l}{\textdagger Simulation did not inspiral, final separation taken at end of simulation.}\\
\end{tabular}
    \caption{Parameters of the inspiral for each simulation. Here $a_0$ is the orbital separation at $t=0$. The times for the start and end of the inspiral (as shown by the circles and triangles in Figure~\ref{fig:evo_vs_time}) are $t_{\rm i}$, and $t_{\rm f}$, respectively, with the corresponding orbital separations given by $a_{\rm i}$ and $a_{\rm f}$. The corresponding period, $P_{\rm f}$, was calculated assuming the primary has the core mass of 0.56~\ms. The time, separation, and timescale at the point at which the inspiral is at its maximum are given by $t_{\rm steep}$, $a_{\rm steep}$ and $\tau_{\rm steep}$, respectively (also shown by the '+' symbols in Figure~\ref{fig:evo_vs_time}). The columns $\langle e_{\rm f -end}\rangle$ and $\langle \dot{a}_{\rm f-end}\rangle$ show the average orbital eccentricity, and rate of orbital decline between $t_{\rm f}$ and the end of the simulation, respectively. Finally, the initial mass transfer rate through $L_1$, measured one year into the simulation is given by $\dot{M}_{\rm {L_1}, 1yr}$.
    }
    \label{tab:CE_summary}
\end{table*}

\subsection{Orbital Evolution}
\label{ssec:orbital_evolution}

\begin{figure*}
    \centering
    \includegraphics[width=\linewidth]{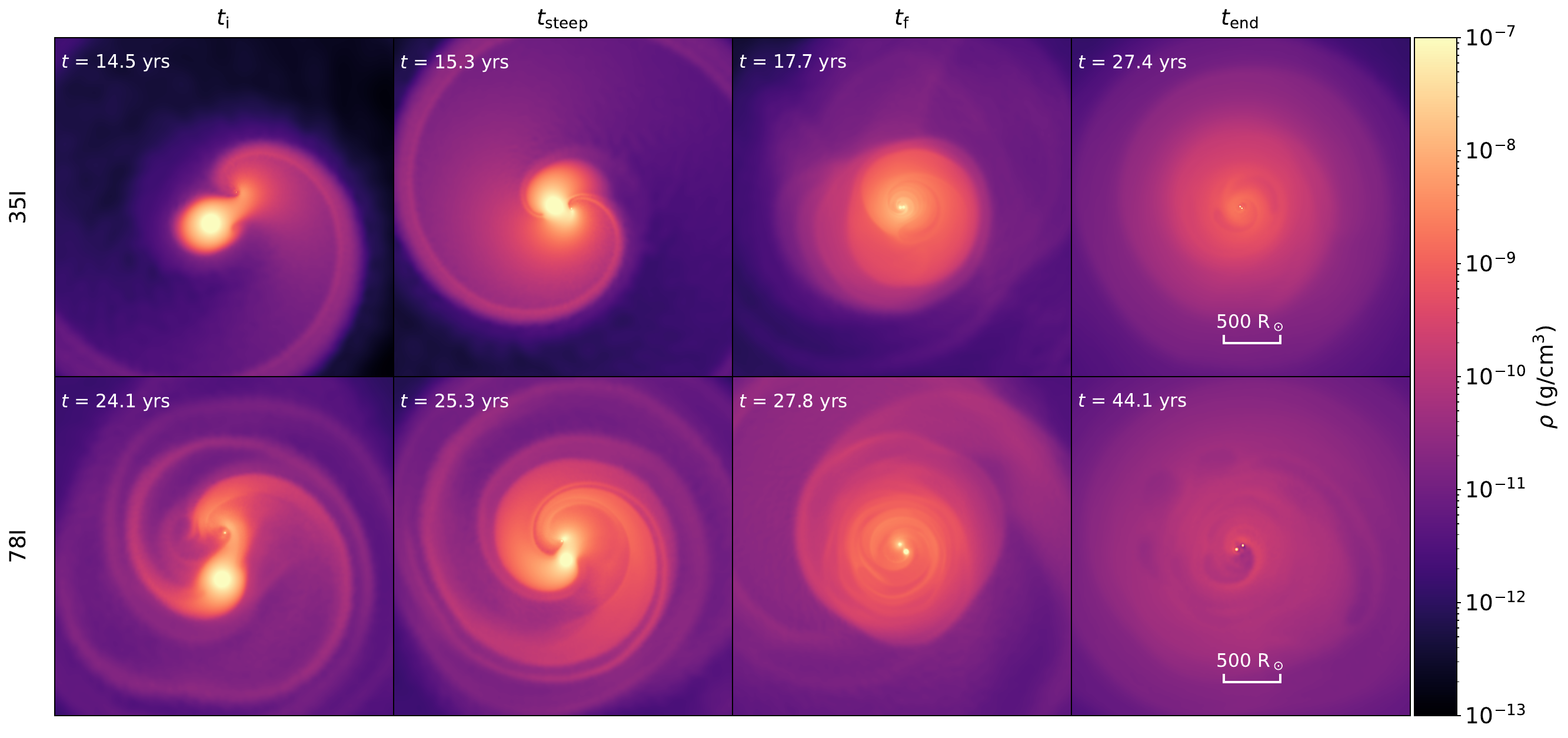}
    \includegraphics[width=\linewidth]{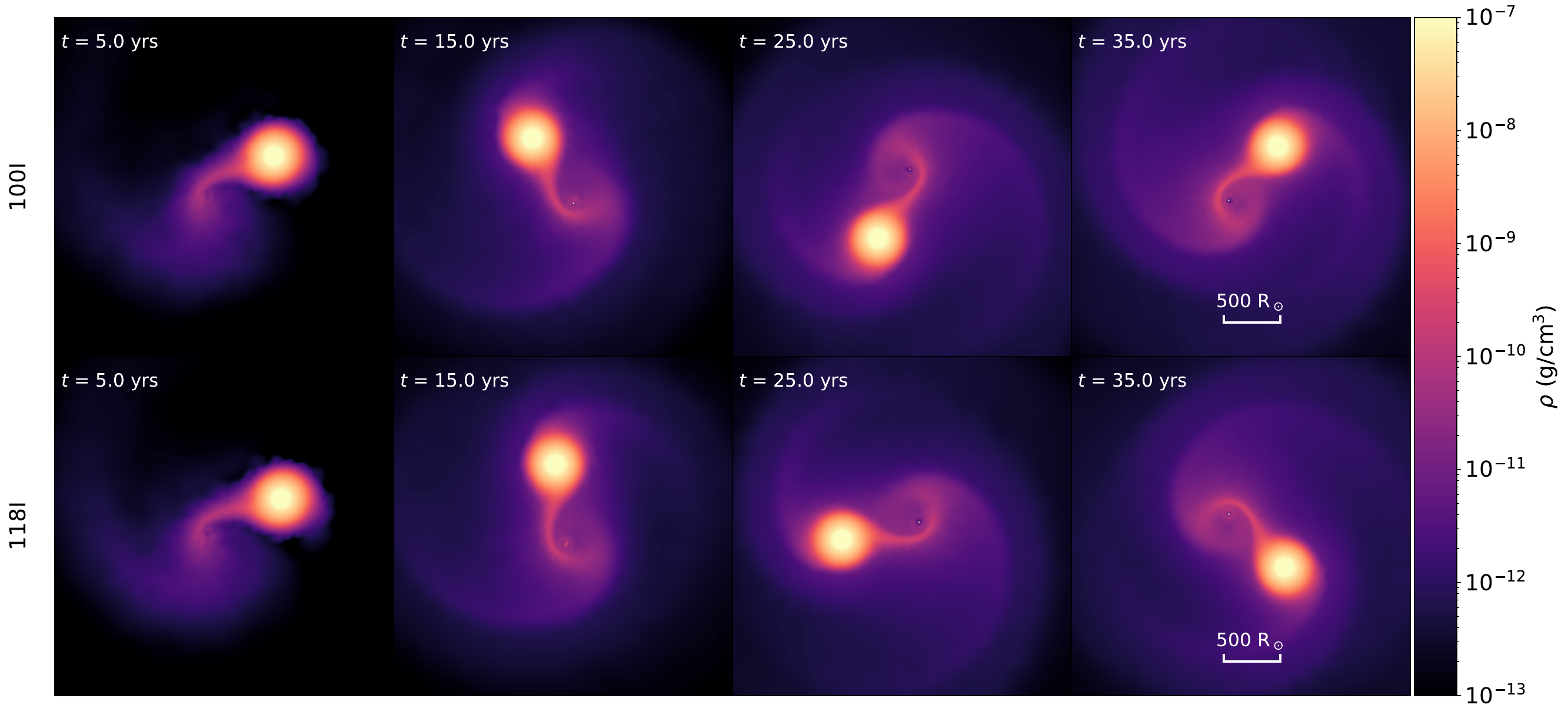}
    \caption{Density slices in the x-y plane for the 35I, 78I (top panel), and the 100I, and 118I simulations (bottom panel). In the top panel, from left to right the columns denote the start of the inspiral ($t_i$), its highest rate of decline ($t_\mathrm{steep}$), the inspiral's conclusion ($t_{f}$), and the final timestep of the simulation  ($t_{end}$). The values are listed in Table.~\ref{tab:CE_summary}. The two high-$q$ simulations shown in the bottom panel did not undergo an inspiral and the snapshots are taken at $t = 5,15,25,35$~yr, respectively. The size of each box is approximately 14~au (3000~\rs) on a side.}
    \label{fig:evo_render_I}
\end{figure*}

Table~\ref{tab:CE_summary} summarises key parameters of the CE interactions. To identify the start and end of the inspiral, $t_{\rm i}$, and $t_{\rm f}$, we adopt the same criterion used by \citet{Reichardt2019}, and \citetalias{Nibbs2025}, namely the first and second time when  $\frac{\dot{a}}{a} = \frac{1}{15} \textrm{max}~|\frac{\dot{a}}{a}|$. Excluding the simulations that do not inspiral (100I and 118I), this definition provides a reasonable characterisation of the beginning and end of the dynamical phase and performs better than in the shallower inspirals considered by \citetalias{Nibbs2025}.

\begin{figure*}
    \centering
    \includegraphics[width=\linewidth]{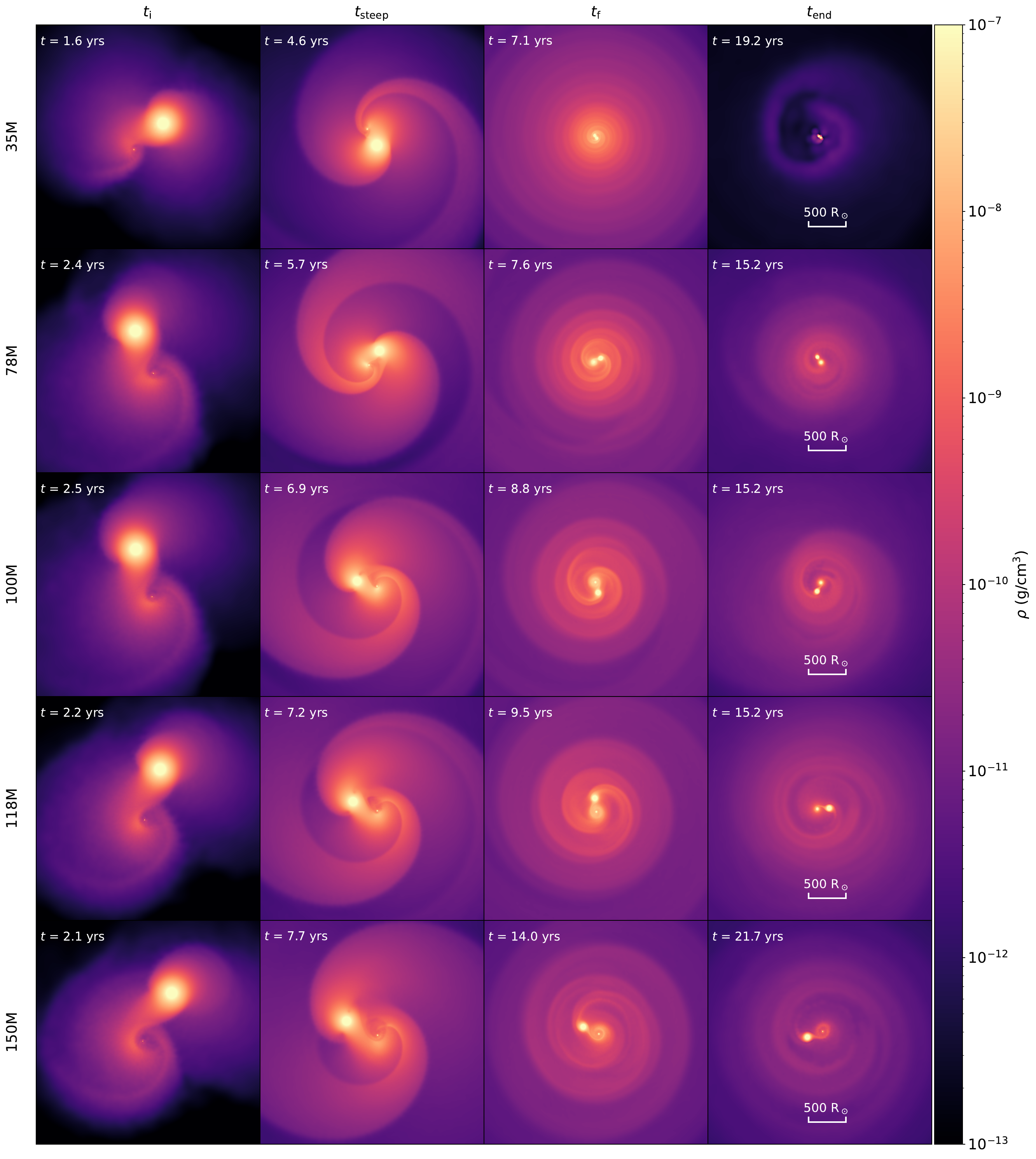}
    \caption{Density slices in the x-y plane for the 35M, 78M, 100M, and 118M simulations from top to bottom, respectively. From left to right the columns denote the start of the inspiral, its highest rate of decline, the inspiral's conclusion, and the final timestep of the simulation. The size of each box is approximately 14~au (3000~\rs) in height and width.
    }
    \label{fig:evo_render_M}
\end{figure*}

Figure~\ref{fig:evo_render_I} and \ref{fig:evo_render_M} display equatorial density slices for all simulations at times $t_i$, $t_{\rm steep}$, $t_f$ and $t_{\rm end}$ (Tables~\ref{tab:simulation_summary} and \ref{tab:CE_summary}). Aside from the two simulations that do not inspiral (100I and 118I; Figure~\ref{fig:evo_render_I} bottom panel) the familiar progression of Roche lobe overflow, common envelope, and rarefaction of the envelope is observed for all simulations. However, it is clear that, particularly for the sequence of four tabulated EoS simulations  (Figure~\ref{fig:evo_render_M}) the larger the value of $q$, the more the two stars  remain distinct, rather than being engulfed into a common stellar envelope.  

\begin{figure*}
\centering
    \includegraphics[width=0.8\linewidth]{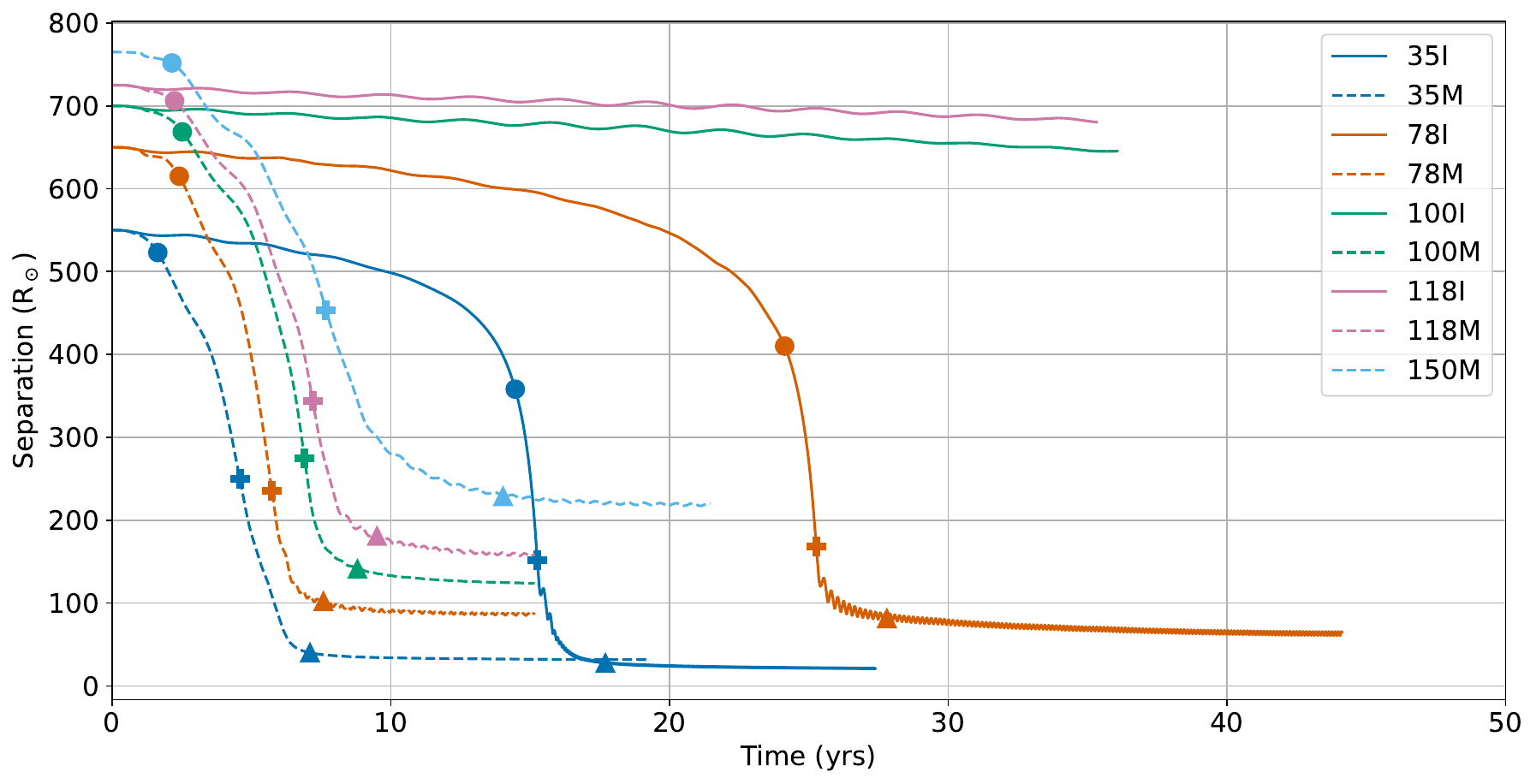}
    \includegraphics[width=0.8\linewidth]{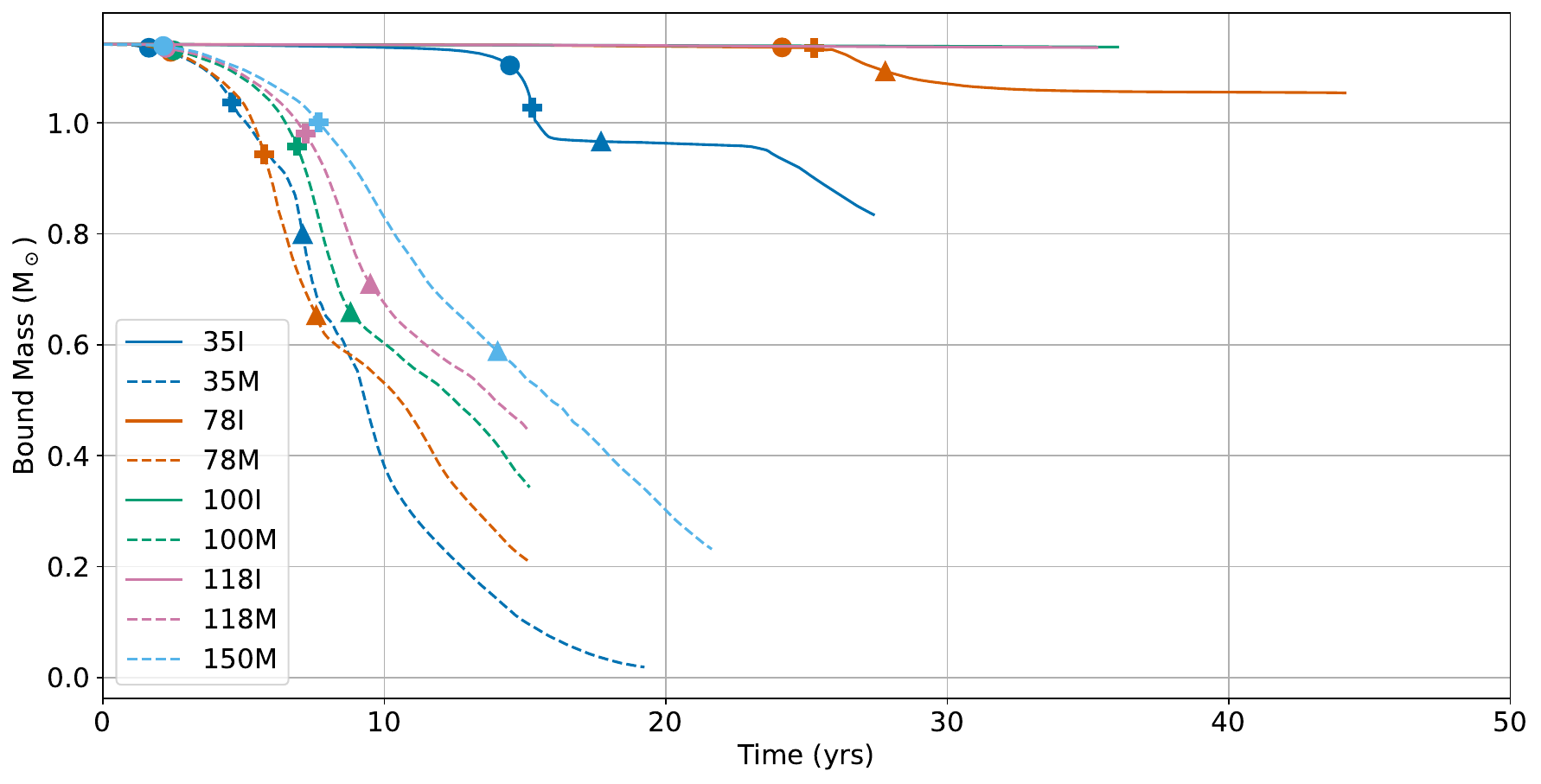}
    \caption{Top panel: Orbital separation between the donor core and companion as a function of time for the 9 simulations listed in Table \ref{tab:CE_summary}. Circles and triangles mark the onset and end of the inspiral respectively, while  '+' symbols indicate the time at which the inspiral rate reaches its maximum magnitude. Simulations 100I and 118I do not undergo an inspiral during the simulated time span.
    Bottom panel: Evolution of the bound mass for the same simulations. Symbols have the same meaning as in the top panel.
    }
    \label{fig:evo_vs_time}
\end{figure*}

In Figure~\ref{fig:evo_vs_time}, top panel, the ideal gas EoS simulations exhibit longer pre-inspiral times for larger $q$, and, above  $q=1.0$, they do not inspiral at all within the simulated time. In contrast, increasing the mass ratio of the tabulated EoS simulations does not significantly extend the duration of the pre-CE mass-transfer phase. In \citetalias{Nibbs2025}, the pre-inspiral timescales increase with $q$ for both EoS simulation sets. There, however, the only simulation that avoids a classic inspiral is the tabulated EoS simulation with the extreme value of $q=1.5$. The inspiral of the tabulated EoS, AGB simulations is not smooth, likely reflecting the nature of the large, diffuse envelope that is being perturbed by the release of recombination energy. 

The final orbital separations, $a_{\rm f}$, range from $28$ to $81$~\rs\ in the two ideal gas simulations that inspiral, and from 40 to 181~\rs\ for the four tabulated EoS simulations. The $q=1.5$ tabulated EoS model yields an even wider final separation of 210~\rs. Overall, $a_\mathrm{f}$ correlates with mass ratio and is slightly larger for tabulated EoS (likely due to the contribution of recombination energy to the envelope expansion reducing the overall density, and with it, the amount of orbital energy extracted from the binary). Each simulation is initialised on a circular, $e=0$ orbit, and at their conclusion, eccentricities remain below $\sim 3.9\times10^{-2}$ (78I).

\begin{figure}
    \centering
    \includegraphics[width=7cm]{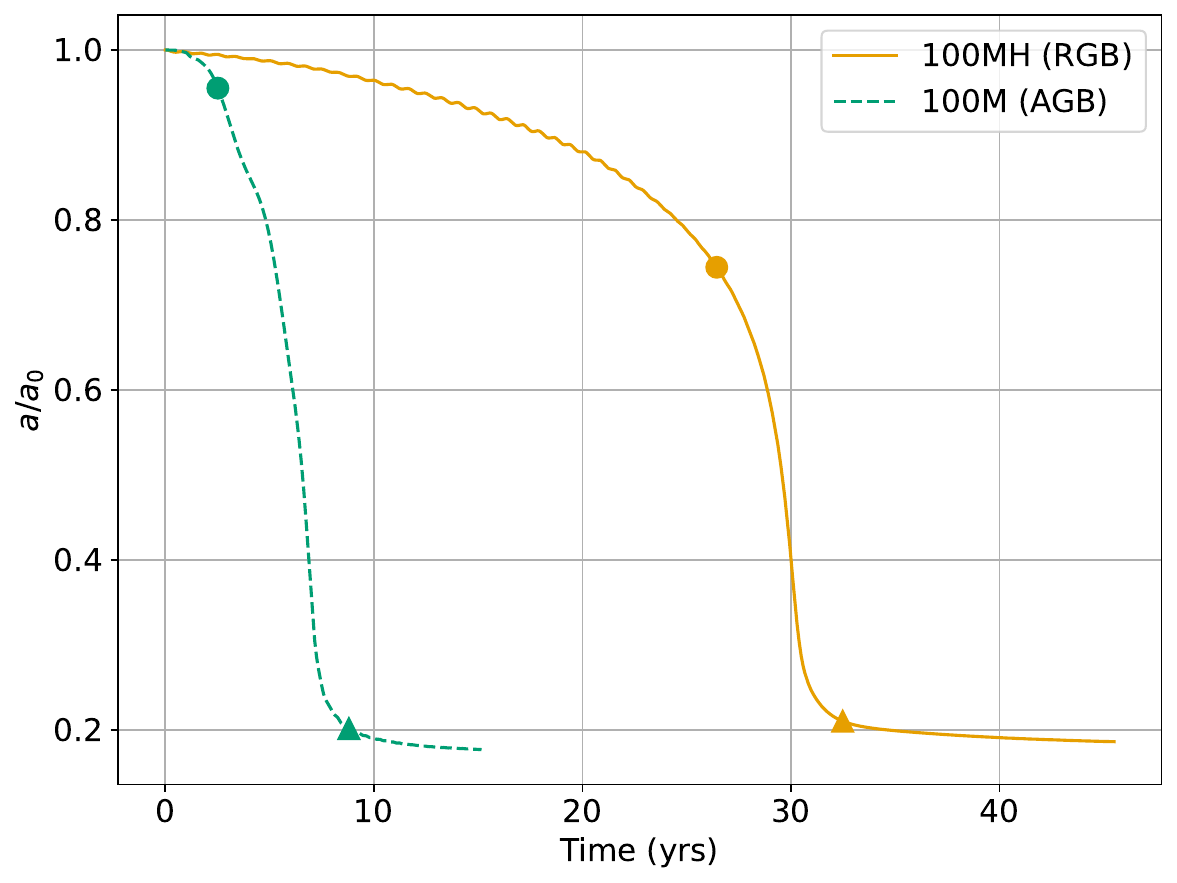}
    \includegraphics[width=7cm]{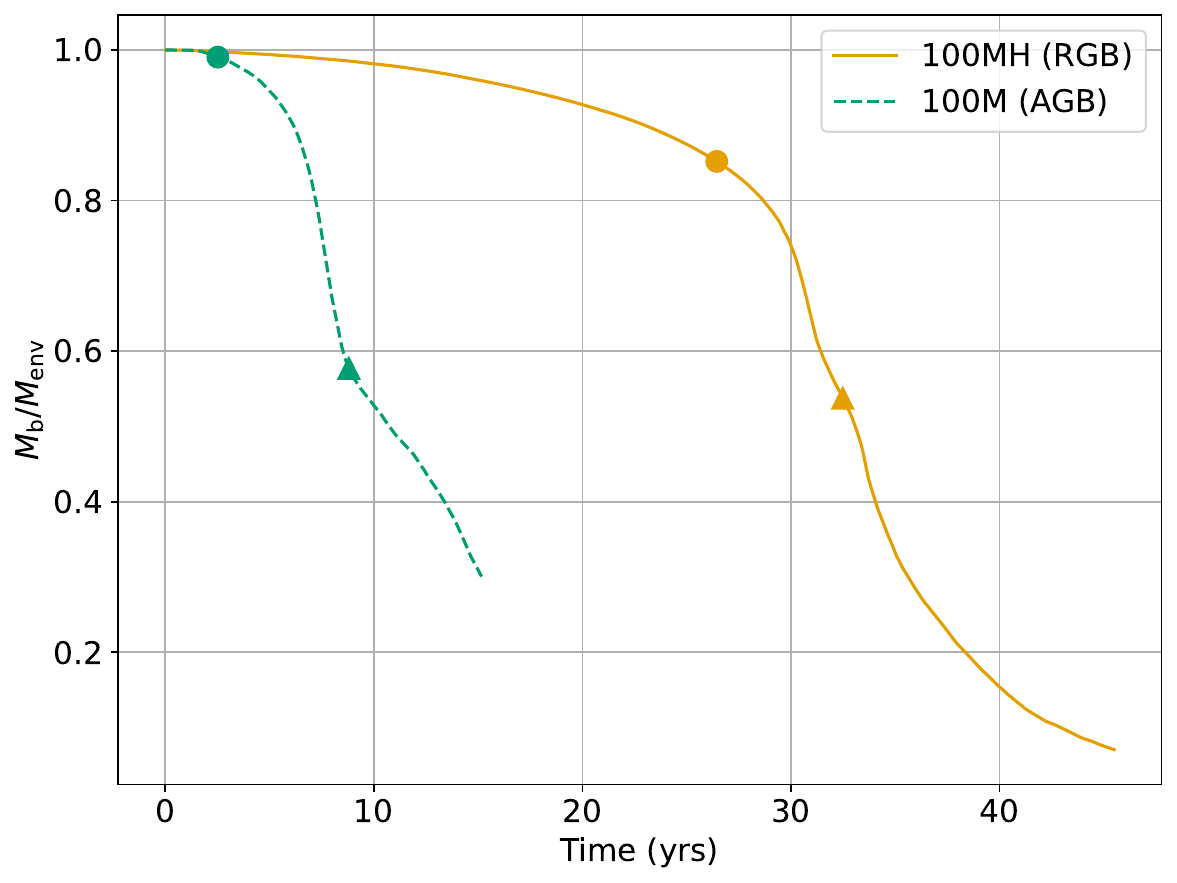}
    \caption{Evolution of the orbital separation (top panel) and bound envelope mass (bottom panel) for the RGB (orange) and AGB (green) simulations with a mass ratio of $q=1$ and a tabulated EoS, taken from \citetalias{Nibbs2025} and this work, respectively. In both panels, quantities are normalised to their initial values, with $a_0=$700 and 235~\rs\ and $M_{\rm env}$=1.14 and 0.49~\ms\ for the AGB and RGB simulations, respectively.}\
    \label{fig:compare_100}
\end{figure}

A larger mass ratio is expected to produce a wider post-CE binary \citep[e.g.,][]{Passy2012}. We ask here whether an extended phase of pre-CE mass transfer, promoted by a larger $q$, can promote and even larger final separation,  having decreased the envelope mass prior to the plunge-in, something that would weaken the subsequent inspiral. We answer this question in Section~\ref{ssec:pre_ce_mass_loss_through_L2_and_L3}.

Comparing the AGB simulations presented here with the RGB simulations of \citetalias{Nibbs2025} (Table~\ref{tab:CE_summary}; table 2 and figure 3 of \citetalias{Nibbs2025}), we find systematically larger post-CE separations for the AGB donors. The final separations span $\sim30 -140$~\rs\ for $q=0.35-1.00$ (AGB), compared to $\sim30 -\sim50$~\rs\ for $q=0.68 - 1.00$ (RGB). As a result, the AGB simulations achieve the lower end of the observed period distribution.

In Figure~\ref{fig:compare_100}, we compare the evolution of the separation (top) and bound mass (bottom) for the 100M simulation (green), and equivalent 100MH simulation (orange) from \citetalias{Nibbs2025}. Despite the substantial differences in donor radius (260~\rs\ versus 90~\rs) and initial orbital separation, the two systems experience similar orbital evolution, each undergoing a reduction of approximately 80\% in separation. Again noting that the inspiral phase is considerably shorter in the AGB case. At $t=1$~yr, we also find that the mass-transfer rate, as measured by the number of particles that have crossed $L_1$ in the AGB simulation, is nearly an order of magnitude larger than in the RGB simulation ($5\times10^{-3}$\ms~yr$^{-1}$ vs $7\times10^{-4}$\ms~yr$^{-1}$). 

Although we do not present a resolution study, previous work has shown that $a_{\rm f}$ is generally well-converged (\citetalias{Nibbs2025}; \citealt{GonzalezBolivar2022,Reichardt2019}). The duration of the pre-inspiral phase, however, is influenced by the mass-transfer rate, which it is itself highly sensitive to the treatment of the stellar atmosphere, which remains poorly resolved in current simulations. Consequently, the duration of the pre-inspiral phase given here is not converged and may be substantially longer in real systems (see also the discussion in  \citet{GonzalezBolivar2022}). 

\subsection{Envelope unbinding}
\label{ssec:envelope_unbinding}

In Figure~\ref{fig:evo_vs_time} (bottom panel) we observe that simulations with a tabulated EoS unbind significantly more envelope mass, both during the inspiral and over the full evolution, due to the release and thermalisation of recombination energy \citep[e.g.,][]{Reichardt2019}. Only the AGB, 35M simulation was evolved long enough to see the end of the mass-unbinding phase, with 98\% of the mass unbound, while the others were terminated early, due to excessive computational times. Judging by the trends and past results \citep[e.g.,][]{Reichardt2020,GonzalezBolivar2022}, however, it is likely that similar fractions of unbound mass would be determined for all AGB, tabulated EoS simulations. All equivalent RGB simulations of \citetalias{Nibbs2025} unbound 93-94\% of their envelope. It is impossible at this stage to judge the exact mass that remains bound to the systems in any simulations, but we estimate that a few percent of the envelope may remain bound.

\begin{figure*}
    \centering
    \includegraphics[width=0.49\linewidth]{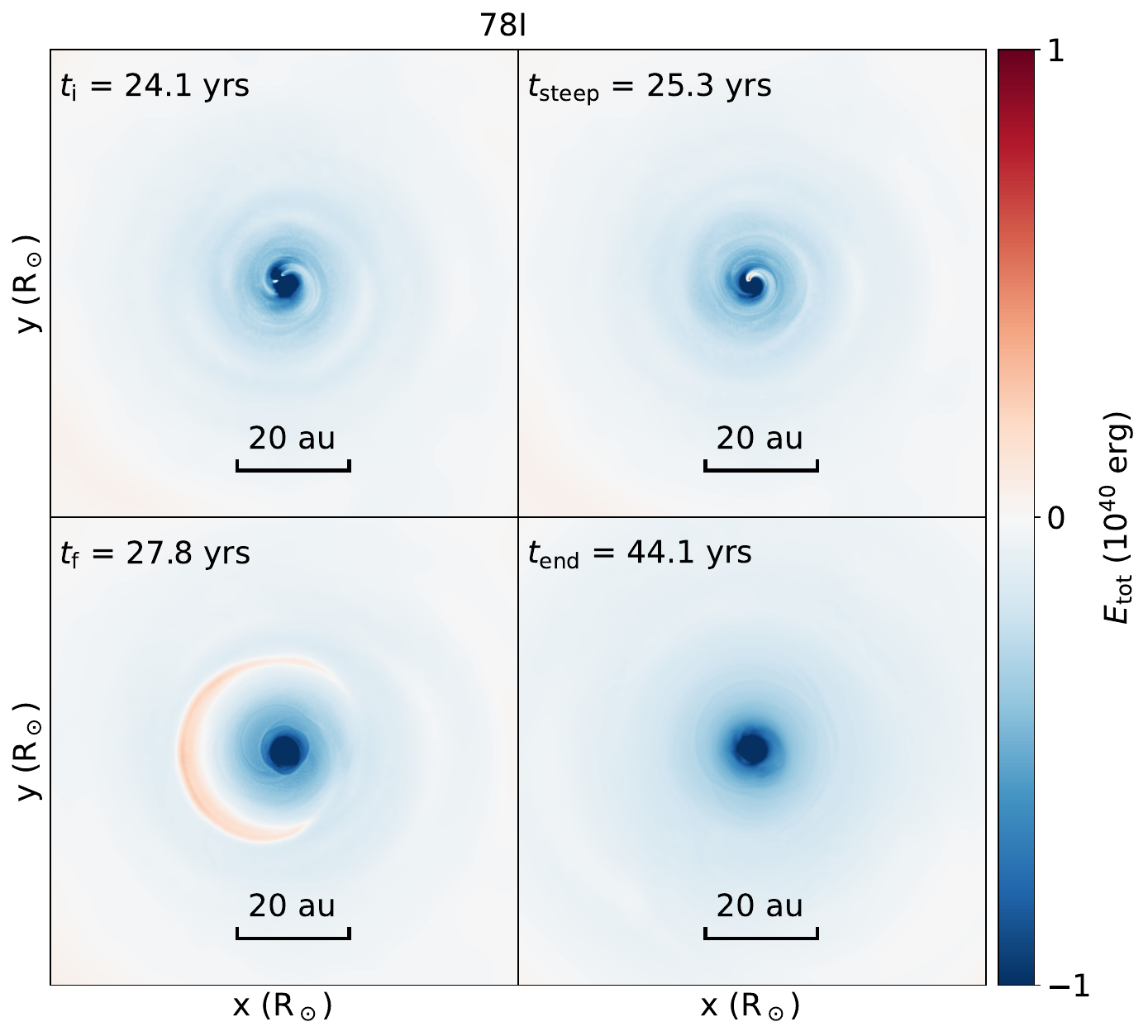}
    \includegraphics[width=0.49\linewidth]{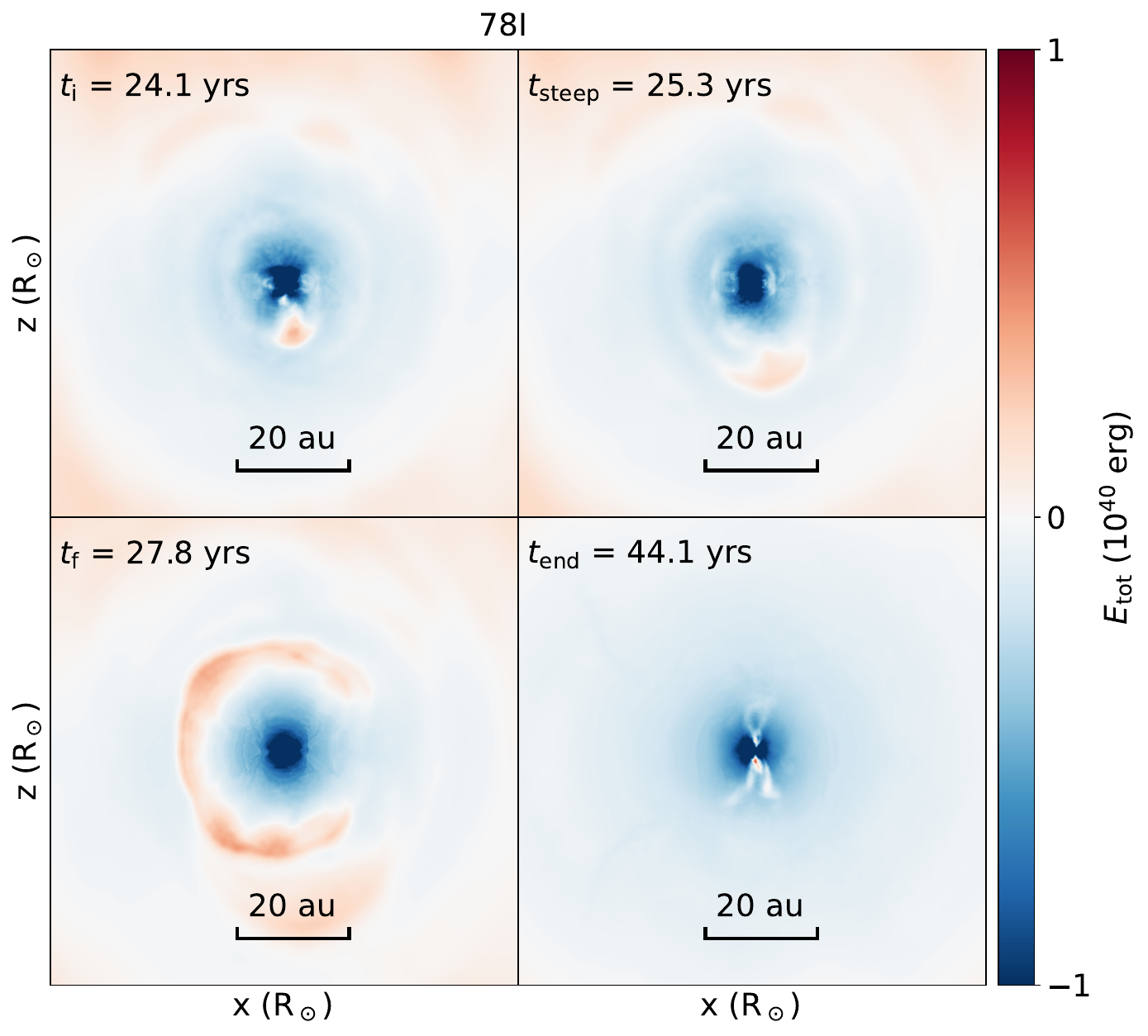}
    \includegraphics[width=0.49\linewidth]{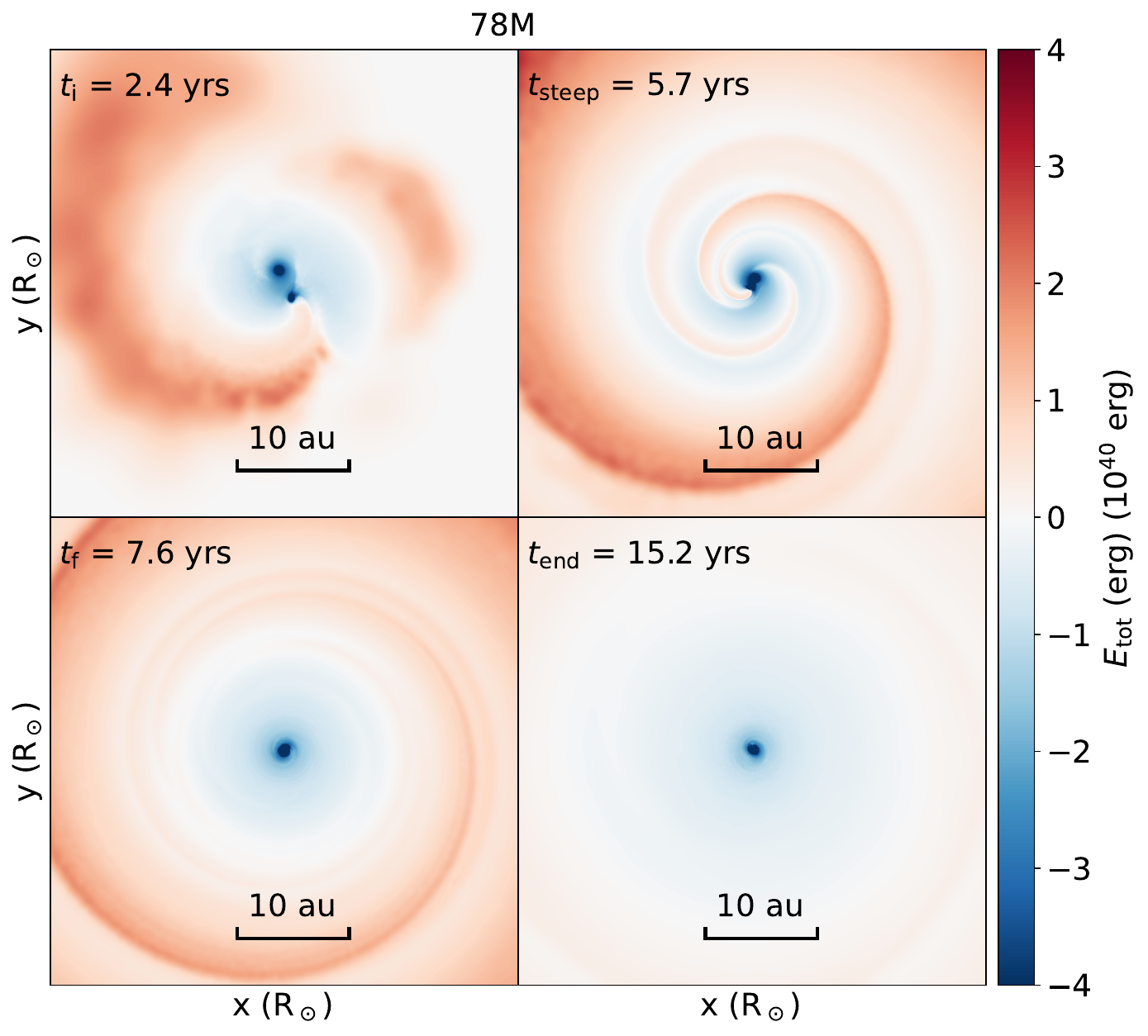}
    \includegraphics[width=0.49\linewidth]{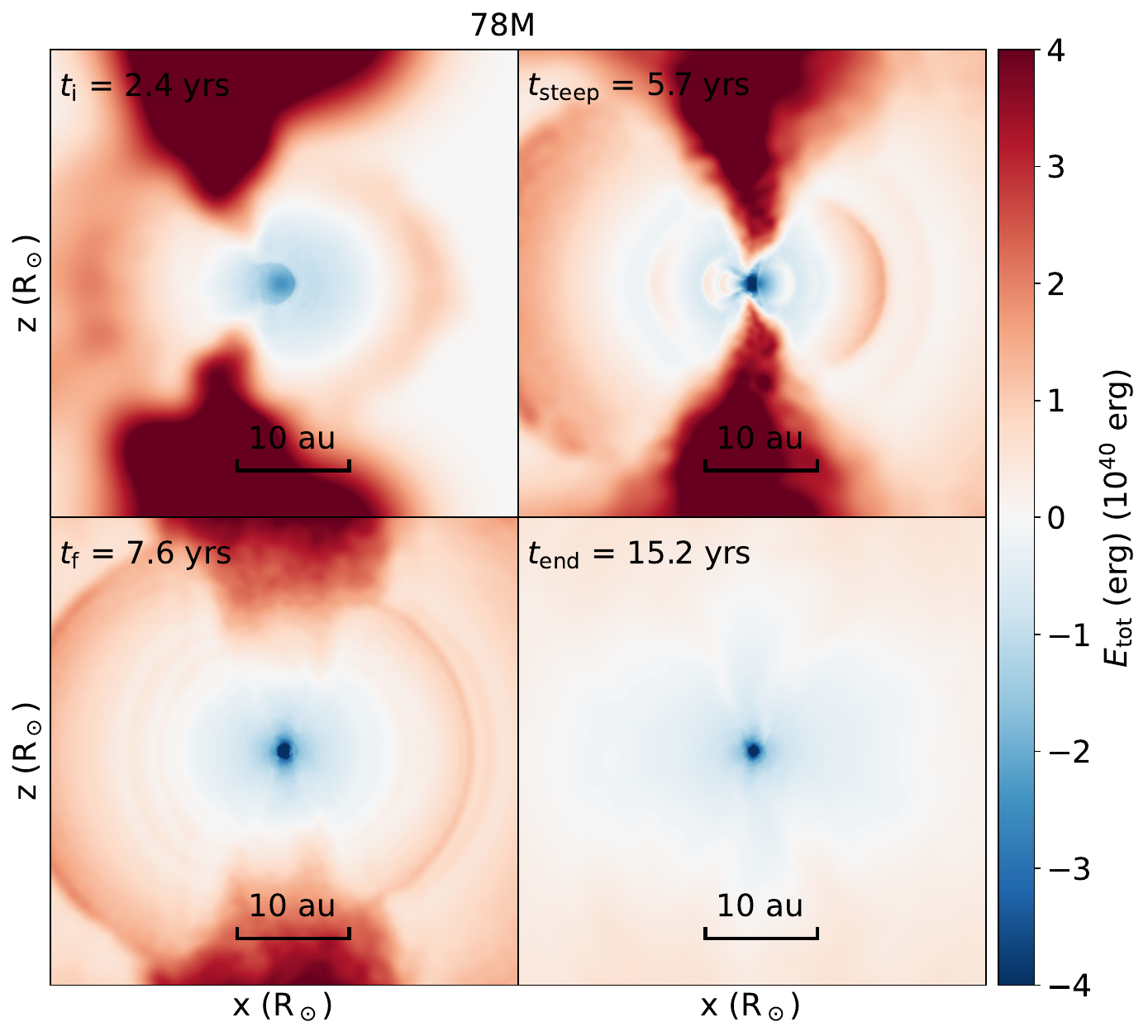}
    \caption{Slices of mechanical energy for the $q=0.78$ simulations. The four panels in each plot proceed chronologically top left to bottom right, showing $t_{\rm i}$, $t_{\rm steep}$, $t_{\rm f}$, and $t_{\rm end}$, respectively (Table~\ref{tab:CE_summary}). The top row shows the 78I simulation, whereas the bottom row shows the 78M simulation. The left and right columns show slices in the x-y and x-z planes, respectively. Note also the colour bar limits increase by a factor of 4 when comparing the top to bottom row, that is, the 78I to the 78M simulation. The size of the top boxes is $\sim$84 au, while the bottom boxes show an area of $\sim$42 au. }
    \label{fig:78_energy_slices}
\end{figure*}

To  characterise further the unbinding process, we present slices of the mechanical energy in the x-y and x-z planes for the 78I and 78M simulations in Figure~\ref{fig:78_energy_slices}. Consistent with the evolution shown in Figure~\ref{fig:evo_vs_time}, only a small fraction of the envelope becomes unbound during the inspiral in 78I. The unbound gas forms an expanding shell with a slightly elongated shape. In the 78M simulation, on the other hand, energetic ejecta are present along the poles with high energy and low mass, while on  the orbital plane, 
mass is ejected from early on in the simulation, mainly along the $L_2$ outflow and out on the outer periphery of the envelope.

\subsection{Bound Mass Distribution and Properties}
\label{ssec:bound_mass_distribution_and_properties}

\begin{figure*}
    \centering
    \includegraphics[width=\textwidth]{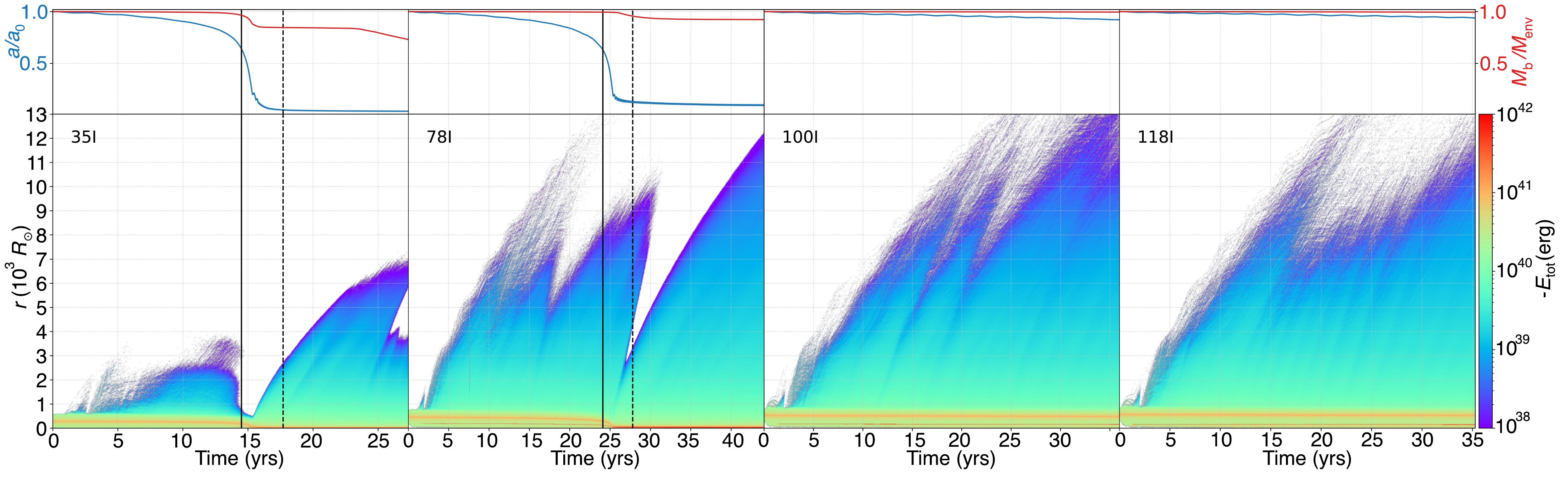}
    \includegraphics[width=\textwidth]{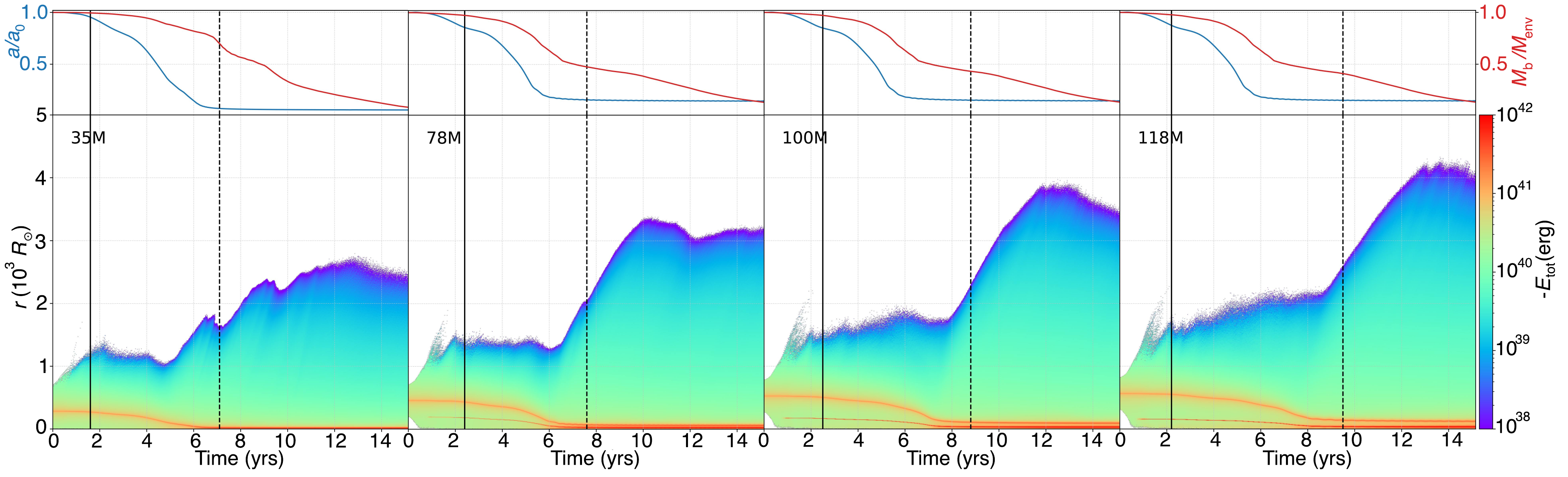}
    \caption{Distribution of bound material ($K+U<0$) in ideal gas simulations  (top row), and tabulated EoS simulations (bottom row - note different y-axis range). Pixels are binned in time and radius with widths of $\approx 10$~d and 5\,\rs, respectively. For each bin, we compute the mean energy per particle 
    of the gas within the corresponding radial shell and plot only bound material. The upper panel of each subplot shows the normalised orbital separation (blue) and bound envelope mass (red). Vertical lines indicate the onset ($t_{\rm i}$, solid) and end ($t_{\rm f}$, dashed) of the inspiral phase, corresponding to the circle and triangle symbols in Figure~5, respectively. These markers are absent for simulations 100I and 118I, which do not undergo an inspiral during the simulated time span.}
    \label{fig:mesa_energy_heatmap}
\end{figure*}

To investigate the dynamics of the bound envelope, we show the evolution of the bound mass distribution as a function of time for all the simulations in Figure~\ref{fig:mesa_energy_heatmap}. Just as in \citetalias{Nibbs2025}, the total specific energy of each SPH particle was computed as the sum of its kinetic, core–gas potential, and gas–gas potential energies (making up the mechanical energy), adding the values for all particles, $k$, contained in a shell $i$ of width $d_i=5$~\rs.

For the ideal gas EoS simulations (Figure~\ref{fig:mesa_energy_heatmap}, top row, first two panels) the deposition of orbital energy is maximum near the companion and most efficient at late times of the inspiral, as can be seen by the deep V-shaped gaps that appear in late inspiral due to the sudden local unbinding of gas at depth.

In contrast, the tabulated EoS simulations preferentially unbind material in regions where recombining occurs, closer to the surface. This is reflected by the overall smaller extent of the bound material in Figure~\ref{fig:mesa_energy_heatmap} (bottom row).

A comparison with the equivalent figures 3 and 4 in \citetalias{Nibbs2025} reveals broadly similar, although not identical, behaviour. In particular, in \citetalias{Nibbs2025}, the RGB simulations with the  tabulated EoS preserve the classic slower, pre-inspiral followed by the plunge-in sequence, resulting in the V-shape typical of sudden deep inspirals causing unbinding near the companion  (figure 4 in \citetalias{Nibbs2025}). On the other hand, here we observe the AGB donor to experience a more gradual inspiral with almost no pre-inspiral, resulting in a gentle expansion,  surface-only unbinding and no V-shape (Figure~\ref{fig:mesa_energy_heatmap}).

This bound material, which extends to radii of  10-50~au, will eventually fall back toward the binary. Provided that it has sufficient angular momentum, it may subsequently contribute to the formation of a circumbinary disc, a possibility we investigate in Section~\ref{subsec:discformation}. 

\subsection{Pre-CE Mass Loss Through $L_2$ and $L_3$}
\label{ssec:pre_ce_mass_loss_through_L2_and_L3}

Mass loss through $L_2$ could play several roles in the evolution of these systems: (i) significant mass lost via $L_2$ prior to the inspiral could lighten the common envelope into which the companion eventually plunges, with the potential of changing the outcome. (ii) Moreover, the fraction of $L_2$ mass loss that remains bound could also be responsible for the formation of the circumbinary disc, if it survived being swept away by ejecta triggered by the subsequent inspiral. (iii) Finally, the efficiency with which the $L_2$ outflow removes angular momentum from the binary contributes to the dynamical reduction of the orbit in all (non-conservative) mass transferring systems, and the quantification of this efficiency, embodied by the parameter $\gamma$, has been of interest in several recent theoretical studies  \citep[e.g.,][]{Macleod2020a,Juarez2026}.

\begin{figure}
    \centering
    \includegraphics[width=\linewidth]{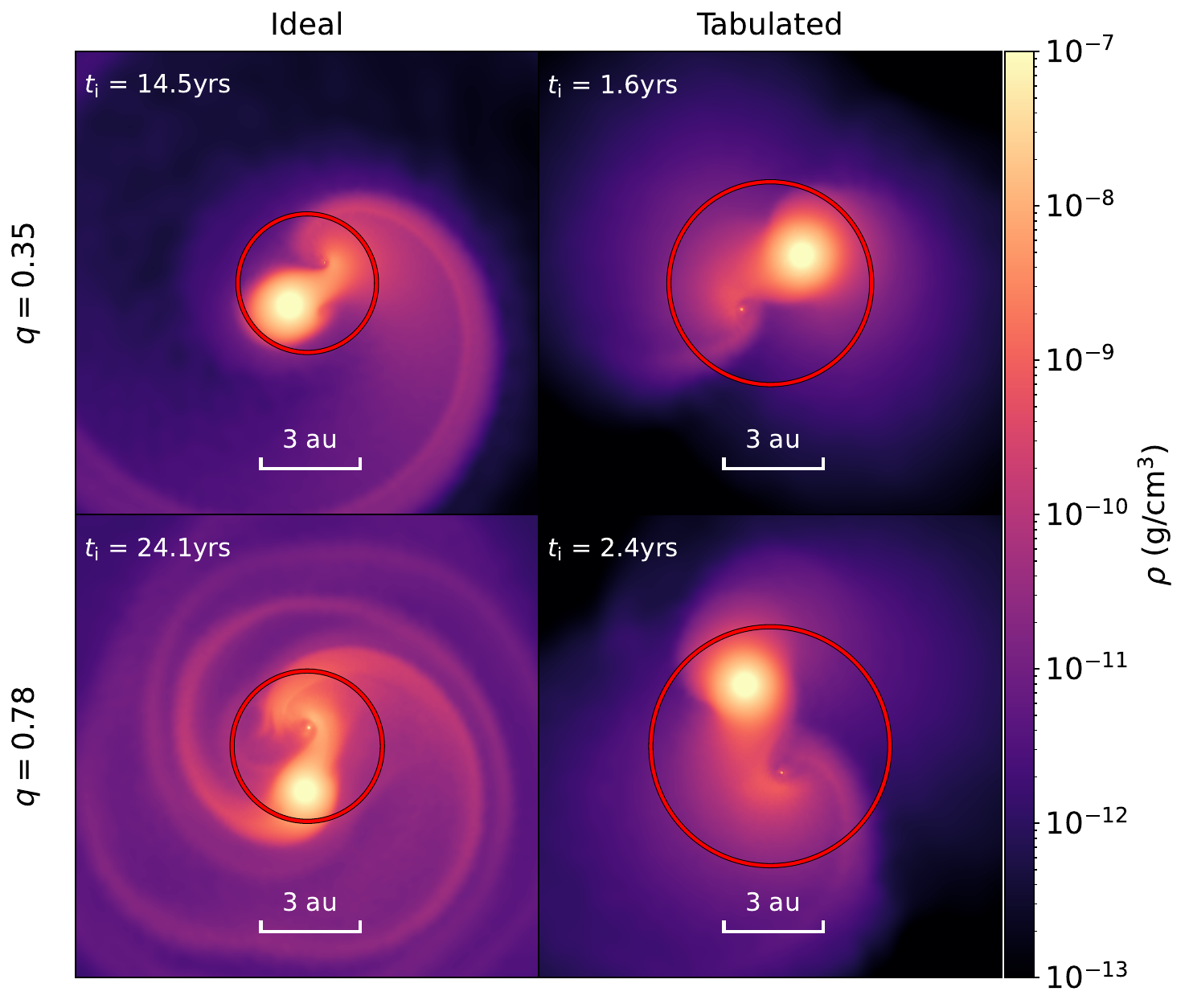}
    \caption{Slices of density in the x-y plane at $t=t_{\rm i}$, the start of the inspiral, for the 35I (top left) 35M (top right), 78I (bottom left) and 78M (bottom right) simulations. The red line in each panel represents the radius of $L_2$, outside of which material is considered pre-inspiral ejecta. Each box is approximately 14 au on a side.}
    \label{fig:L2_circle_render}
\end{figure}

\begin{figure*}
    \centering
    \includegraphics[width=0.4\linewidth]{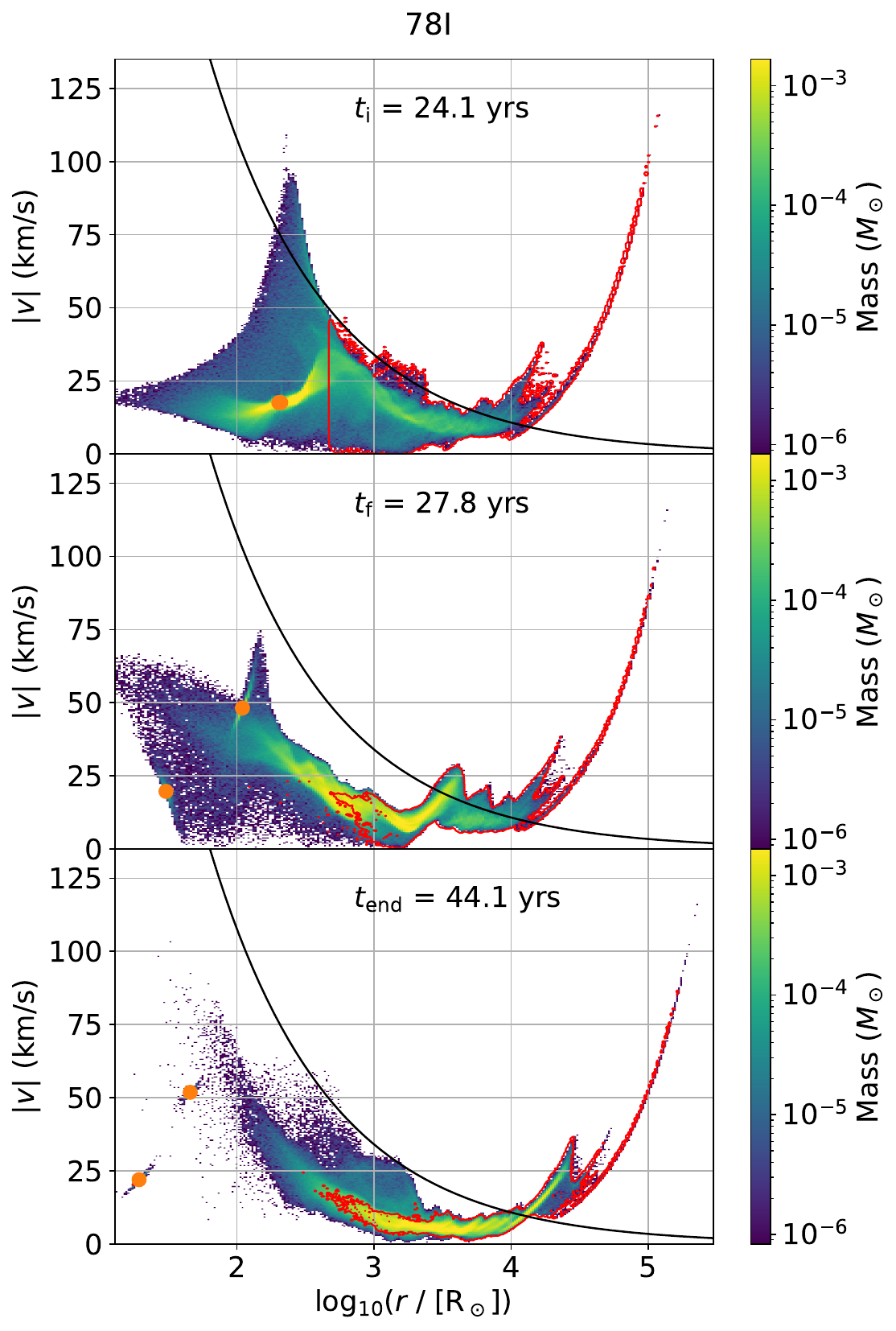}
    \includegraphics[width=0.4\linewidth]{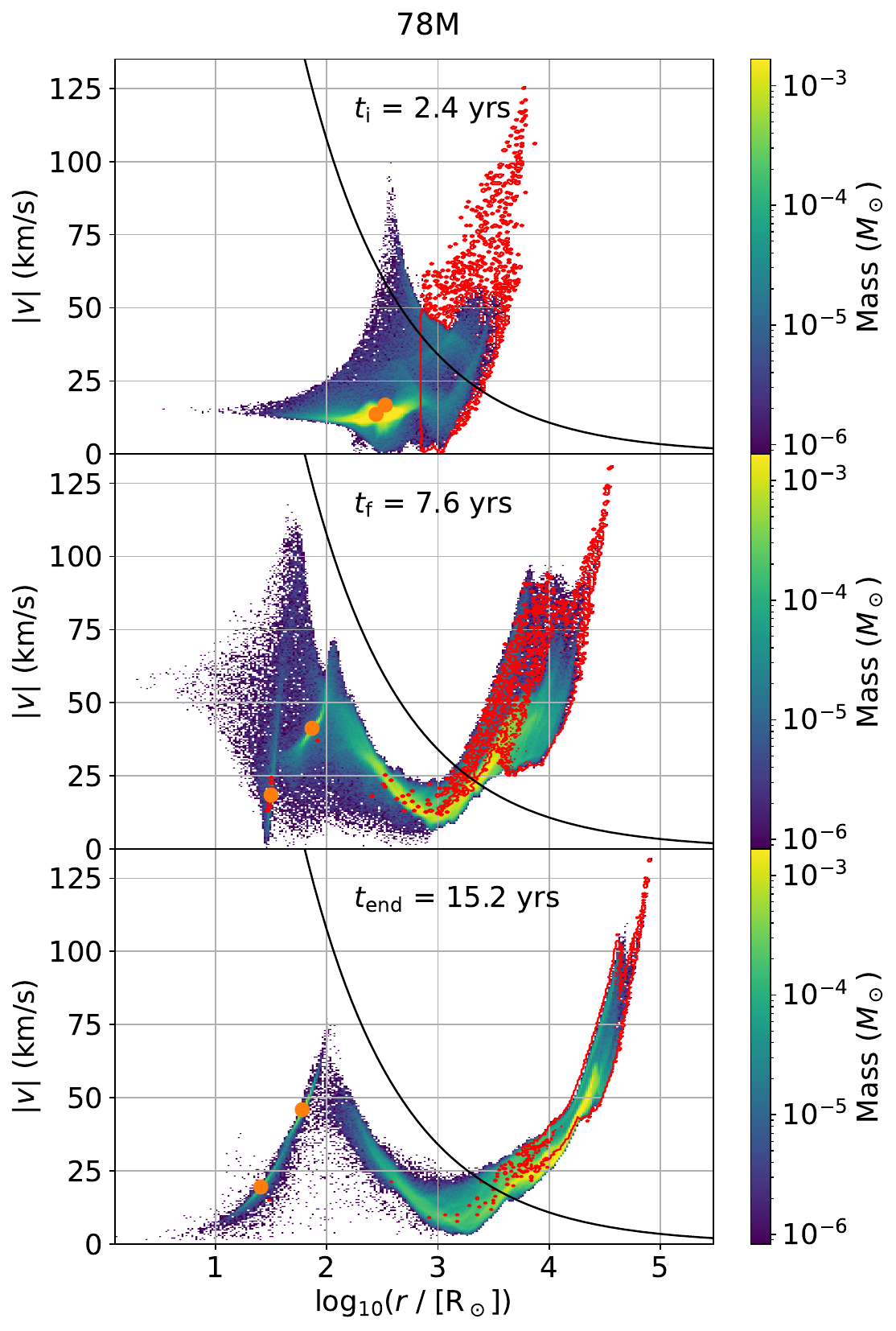}
    \caption{Velocity of each SPH particle as a function of distance from the binary's centre of mass for the 78I (left) and 78M (right) simulations. The times shown in chronological order from top to bottom are the $t_{\rm i}$, $t_{\rm f}$, and $t_{\rm end}$ (Table~\ref{tab:CE_summary}). Black curve: escape velocity assuming the central mass to be the primary star {\it and} the companion --- providing an upper limit value. The particles bounded by the red contour are those located outside the radius of $L_2$ at the onset of the inspiral, $t_{\rm i}$, as shown in Figure~\ref{fig:L2_circle_render}. To display the mass distribution, we construct a 2D histogram of mass using 300 logarithmically spaced radial bins between the minimum particle radius (or $1\,R_\odot$, whichever is larger) and $3\times10^5\,R_\odot$, and 300 linearly spaced velocity bins between 0 and 135 km s$^{-1}$, where the mass shown is the mass per bin. Orange circles: location of the core and companion.}
    \label{fig:vesc_78}
\end{figure*}

To quantify the amount of mass lost prior to the onset of the inspiral and its dynamics, we estimate the location of $L_2$  using the prescription of \citet[][see Paper~I, equation 2]{Misra2020}. As also explained in \citetalias{Nibbs2025},  we have modified that prescription so as to measure the location of $L_2$ from the centre of mass of the system, and not from the centre of the donor\footnote{For $q>1$ the locations of the Lagrange points swap such that $L_2$ is behind our donor star. Here, however, we retain the nomenclature that $L_2$ is the Lagrange point behind the companion, even when the companion is more massive than the primary.}.
We then count each SPH particle that moves beyond this radius at the onset of the inspiral, $t_{\rm i}$. This method of counting also includes material ejected through $L_3$. We plot the $L_2$ radius in Figure~\ref{fig:L2_circle_render} for four simulations, as a red dashed line at time $t=t_{\rm i}$. The material outside of this radius is called ``the $L_2$ gas'' in Table~\ref{tab:L2_ejecta_summary}, where we provide a summary of the bound or unbound properties for this gas as well as for the entire envelope. Further, in Figure~\ref{fig:vesc_78} we visualise that $L_2$ gas (for representative simulations 78I and 78M), by showing the velocity as a function of distance from the centre of mass, alongside the value of the escape speed for all SPH particles, where we enclose the $L_2$ gas within the red contour. 

\begin{table*}[]
    \centering
    \begin{tabular}{cccccccccccc}
    \toprule
  (1)& (2)& (3)& (4)& (5)& (6)& (7)& (8)& (9)& (10)&(11)&(12)\\
 Sim. & $r_{\rm L_2}$ & $t_{\rm i}$ & $M_{>L_2, \textrm{i}}$ & $M_{>L_2, \textrm{UB}, \textrm{i}}$ & {$M_{>L_2, \textrm{UB}, \textrm{i}}$} & $t_{\rm f}$& $M_{>L_2, \textrm{UB}, \textrm{f}}$ & {$M_{>L_2, \textrm{UB}, \textrm{f}}$} & $M_{\rm Tot, UB, i}$ & $M_{\rm Tot, UB, f}$ &$M_{\rm Tot, UB, end}$\\

   Name &  (\rs) &   (yr) &  (\% $M_{\rm env}$) &  (\% $M_{\rm env}$) &  (\% $M_{>L_2, \textrm{i}}$)&  (yr) &  (\% $M_{\rm env}$)  &  (\% $M_{>L_2, \textrm{i}}$) &  (\% $M_{\rm env}$) &  (\% $M_{\rm env}$) &(\% $M_{\rm env}$) \\
\midrule
   35I &  449 &  14.5 &   4.8&     3.0&        63&  17.7&      3.5&         73&     3.4 &     15 &  27 \\
   35M &  657 &   1.6 &   1.5&     0.5&        33&   7.1 &      1.3&         87&     0.5 &     30 &  98\\
   78I &  487 &  24.1 &  20&     0.5&         2.5&  27.8&      0.6 &          3&     0.5 &      4.3 &   7.7 \\
   78M &  743 &   2.4 &   2.7 &     1.0 &        37&   7.6 &      2.4 &         89&     1.2 &     43 &  $<$82 \\
  100I*&  772 &     $-$ &   4.5 &     0.5 &        11&     $-$ &      $-$ &         $-$ &     $-$ &      $-$ &   0.5 \\
  100M &  800 &   2.5 &   2.8&     0.8&        29&   8.8 &      2.3&         82&     0.9 &     42 &  $<$70 \\
  118I*&  826 &     $-$ &   4.2 &     0.5 &        12&     $-$ &      $-$ &        $ -$ &     $- $&      $-$ &   0.5 \\
  118M &  859 &   2.2 &   2.2&     0.4&        18&   9.5 &      1.8&         82&     0.6 &     38 &  $<$61 \\
150M &  929 &   2.1 &   2.1 &     0.2 &          10 &  14.0 &      1.7 &           81 &     0.3 &      0.0 &  $<$22\\

\bottomrule
\multicolumn{12}{l}{*Did not undergo an inspiral, quantities are calculated at the end of the simulations $t=t_{\rm end} = 36.1$ and 35.3 years, for 100I and 118I, respectively.}
\end{tabular}
    \caption{Properties of the mass lost through $L_2$ by the start of inspiral, $t_i$. Columns are as follows: (2): $r_{\rm L2}$ --- distance of $L_2$ from the centre of mass at the onset of the dynamical inspiral, $t_{\rm i}$ (3); 
    (4): $M_{>L_2, \textrm{i}}$ --- percentage of the envelope mass outside  $r_{\rm L_2}$ at time $t_{\rm i}$; (5) $M_{>L_2, \textrm{UB, i}}$ --- unbound mass outside of $r_{\rm L2}$ at time $t_{\rm i}$ as a percentage of envelope mass, and of the mass exterior to $L_2$ (6), respectively; (7): $t_{\rm f}$ --- end of the CE inspiral (triangles in Figure~\ref{fig:evo_vs_time});  $M_{>L_2,\textrm{UB, f}}$ --- mass outside $r_{\rm L2}$ at $t_i$ that is unbound at $t_f$, as a percentage of envelope mass (8), and as a percentage of mass exterior to $L_2$ (9); (10): $M_{\textrm{Tot, UB, i}}$, and (11) $M_{\rm Tot,UB,f}$ --- total unbound mass in the simulation at $t_i$ and $t_f$, respectively (circles, and triangles in Figures~\ref{fig:evo_vs_time}, respectively); (12) $M_{\rm Tot,UB,end}$ --- total envelope mass unbound at the end of the simulation. For these evaluations, $M_{\rm env} = 1.14$~\ms.
    \label{tab:L2_ejecta_summary}}
\end{table*}

Inspecting Table~\ref{tab:L2_ejecta_summary} and table~3 in Paper~I, we can answer question (i) posed above, in the negative: higher mass ratios do not systematically lead to more mass loss via $L_2$ prior to inspiral. The wider final separation, $a_f$, attained in higher $q$ simulations is due to the larger injection of orbital energy by the more massive companion, and {\it not} by a lighter envelope at $t_i$. The RGB, Paper~I, simulation sequences $q = 0.68, 0.85$ and 1.0 (both EoS), {\it and} the AGB simulation sequence presented here, $q=0.35, 0.78, 1,0$ and 1.18 (but only for the tabulated EoS) all remove small amounts of envelope by the time of inspiral with no trends with $q$. The ideal gas EoS, AGB simulation sequence  {\it mostly} agrees with this conclusion, although simulation 78I does remove an anomalously high (20\%) of the envelope, four times larger than 35I (5\%), while simulations 100I and 118I, ejected $< 5$\% of the envelope, but do not inspiral so they cannot be fairly compared. 

In response to point (ii) above, a few percent of the envelope leaves the $L_2$ and remains bound to the system (Table~\ref{tab:L2_ejecta_summary}). This could form a circumbinary disk. However, when the CE inspiral takes place, mass loss from the inspiral sweeps away the $L_2$ disk. All  simulations here and in Paper~I, {\it except for  three}, tend to sweep and unbind material by simulation's end, independently of how much $L_2$ outflow remains bound at $t_i$ (see Figure~\ref{fig:vesc_78}; values are reported in Table~\ref{tab:L2_ejecta_summary}).  The three simulations that avoid a CE inspiral are the RGB simulations with $q=1.5$, that have only a gentle inspiral, but unbind most-to-all of their $L_2$ gas. The two AGB simulations with $q=1.0$ and 1.18 and ideal gas EoS do not inspiral and the 10\% of the envelope that leaves through $L_2$ remains almost entirely bound.

As found in all past simulations \citep[e.g.,][]{Nandez2016,Ohlmann2016,Reichardt2020}, all our tabulated EoS simulations unbind the envelope (if they are followed to the end of the inspiral). Taking the 35M simulation from this paper as an example, the CE ejecta sweeps all $L_2$ gas and becomes almost entirely unbound. The $\sim$2\% that is not unbound could return to the system forming a fall-back disk, a topic we explore in Section~\ref{subsec:discformation}.

Finally, in response to point (iii) above, Table~\ref{tab:MJ_summary} presents the mass and angular momentum transported beyond $L_2$ and $L_3$ by the time of the onset of the inspiral ($t=t_{\rm i}$). Following \citetalias{Nibbs2025}, we use the $\gamma_{\rm loss}$  parameter \citep{Nelemans2000}, which relates the specific angular momentum carried away by the outflow, $\Delta J/\Delta M$, to that of the binary, $J_{\rm bin}/M_{\rm bin}$, or, equivalently, the fraction of binary angular momentum lost, $\Delta J / J_{\rm bin}$, to the corresponding mass lost, $\Delta M/M_{\rm tot}$:

\begin{equation}
    \frac{\Delta J}{J_{\rm bin}} = \gamma_{\rm loss}\frac{\Delta M}{M_{\rm tot}}.
\end{equation}
Higher values of $\gamma_{\rm loss}$ correspond to more efficient removal of orbital angular momentum per unit mass lost, and can lead to a stronger orbital response (see equation 2 of \citealt{MacLeod2018a}). 

\begin{figure*}
    \centering
    \includegraphics[width=0.9\linewidth]{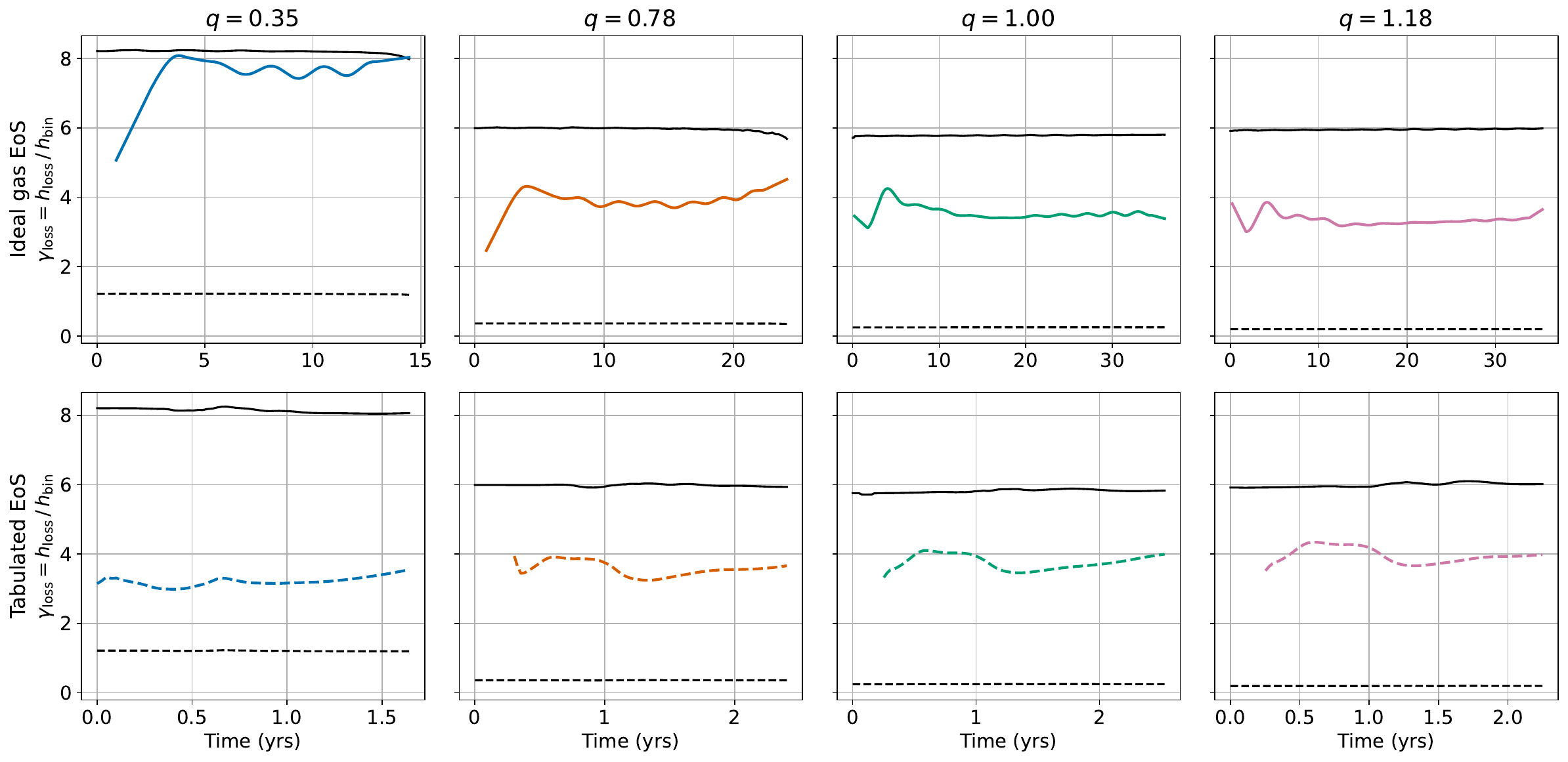}
    \caption{Evolution of $\gamma_{\rm loss}$ as a function of pre-inspiral time for all simulations. Solid and dashed black lines show $\gamma_{\rm L_2}$ and $\gamma_{\rm acc}$, respectively. The top and bottom rows show the ideal gas (I) and tabulated EoS (M) simulations, while columns are ordered by mass ratio. Early-time values, especially in the ideal gas simulations, are noisy and should be interpreted with caution, as they are based on only a few particles carrying a large fraction of the angular momentum loss.}
    \label{fig:gamma_loss_vt}
\end{figure*}

For the RGB simulations of \citetalias{Nibbs2025}, we typically found $\gamma_{\rm loss}\approx4$, with values closer to 2 for the $q=1.50$ simulations that did not undergo a classical CE inspiral\footnote{\citetalias{Nibbs2025} used a slightly different method to calculate $\gamma$; applying the method adopted here to those simulations gives somewhat larger values, $\gamma\approx4-5$, without altering the conclusions of that work (for completeness we have recalculated the values of \citetalias{Nibbs2025} and we present them in \ref{sec:appen3}).}. For the AGB simulations, $\gamma_{\rm loss}\approx3$ in most cases (Figure~\ref{fig:gamma_loss_vt}). In Paper~I and for the ideal gas EoS simulations here, there is a slight but clear trend of decreasing $\gamma_{\rm loss}$ with increasing $q$, while for the AGB tabulated EoS simulations here that trend is weak but is inverted.

\begin{table*}[]
    \centering
    \begin{tabular}{cccccccc}
    \hline
         (1) &   (2)& (3) & (4)  & (5)  &(6)& (7)&(8)\\
         Sim.& 
     $M_{\rm i}$& $J_{\rm i}$& $\langle \gamma_{\rm loss}\rangle$& $\langle \gamma_{\rm L_2}\rangle$&$\langle \gamma_{\rm acc}\rangle$& $h_{\rm bin, 0}$&$J_{\rm bin,0}$\\
 Name& (\% total)& (\% total)& &  && ($10^{19}$ cm$^2$s$^{-1}$) &($10^{52}$g cm$^2$s$^{-1}$)\\
 \hline
 35I& 2.4& 15.4& 7.6& 8.2 &1.2& 2.1&9.5\\
 35M& 0.8& 2.6& 3.2& 8.2&1.2& 2.1&9.5\\
 78I& 7.3& 26& 3.9& 6.0 &0.4& 3.3&20\\
 78M& 1.0& 3.5& 3.6& 6.0 &0.4& 3.3&20\\
 100I~\textdagger & 1.5& 3.9& 3.5& 5.8&0.2& 3.7&25\\
 100M& 0.9& 3.5& 3.8& 5.8&0.2& 3.7&25\\
 118I~\textdagger & 1.3& 5.1& 3.3& 6.0&0.2& 3.9&28\\
 118M& 0.7& 2.7& 4.0& 6.0&0.2& 3.9&28\\
 150M& 0.6& 2.3& 4.3& 6.4& 0.1& 4.1&35\\
 \hline
 \multicolumn{8}{l}{\textdagger~ Did not go through a CE inspiral, value is instead taken at the end of the simulation.} \\ 
 \end{tabular}
    \caption{Properties of the material located beyond $r_{L_2}$ at the onset of the inspiral, $t_i$. Note that this also includes some of the material lost through $r_{L_3}$. Columns are as follows. (2), $M_{\rm i}$: mass exterior to $L_2/L_3$ expressed as a percentage of the total system mass at $t_i$; (3), $J_i$: angular momentum outside $L_2$ at $t_i$, as a percentage of the system's total angular momentum at the same epoch.  $\gamma_{\rm loss} = h_{\rm loss}/h_{\rm bin}$,  $\gamma_{\rm L_2} = h_{\rm L_2}/h_{\rm bin}$ and $\gamma_{\rm acc} = h_{\rm acc}/h_{\rm bin}$ are the specific angular momenta of the mass lost from the binary, of $L_2$ and of $M_2$, respectively, all relative to the binary's initial specific angular momentum. So (4-6), are the respective averages from the start of the simulation to $t_i$. (7) and (8), $h_{\rm bin,0}$ and $J_{\rm bin,0}$: the binary's initial specific and total angular momenta, respectively. }
    \label{tab:MJ_summary}
\end{table*}

\subsection{Fall-back Material and Circumbinary Disc Formation}\label{subsec:discformation}

We now consider the size and mass of a possible fall-back disc. We assume the bound ejecta follows ballistic trajectories, conserving angular momentum. At any point along this orbit, the approximate time for the particle to return to pericentre will be on the order of half the orbital period. Thus, on average, the fall back time for a particle, or gas parcel  is
\begin{equation}
    t_{\rm fb} = \pi\sqrt{\frac{a^3}{GM}},
    \label{eq:t_fb}
\end{equation}
where $M$ is the total central mass (primary core and companion), $G$ is the gravitational constant and $a$ is the semi-major axis of the orbit (derived from each SPH particle velocity vector, distance from the centre of mass and central mass). This bound ejecta will circularise at the circularisation radius,  
\begin{equation}
    R_{\rm c} = \frac{h^2}{GM},
    \label{eq:Rc}
\end{equation}
where $h$ is the specific angular momentum of the ejecta.

The expected timescale over which this material will circularise will be of the order of the orbital period (\citealt{Rossi2021}\footnote{Even without general relativistic precession driving circularisation in the work of \citet{Rossi2021}, extended fallback around a binary operates in Rossi’s fast regime, because the 3D distribution of infalling gas, combined with the binary's non-Keplerian quadrupole potential, forces trajectories from different azimuths and inclinations to intersect directly near $R_{\rm c}$ on their very first pericenter passage, dissipating their radial kinetic energy within a few local dynamical times.}) at this new semi-major axis given by $\tau_{\rm circ} \sim  2\pi \sqrt{R_c^{3}/GM}$. The disc is then expected to evolve and spread over time due to viscous interactions causing a redistribution of the angular momentum on a timescale given by
\begin{equation}
    \tau_{\rm visc}=\frac{R_{\rm c}^2}{\nu},
    \label{eq:tVisc}
\end{equation}
where $\nu = \alpha c_{\rm s}^2\sqrt{R_{\rm c}^3/GM}$ is the gas' kinematic viscosity,  with $c_s$ the sound speed and $\alpha$ the disc's viscosity parameter \citep{Shakura1973,artymowicz_dynamics_1994}.

In Table~\ref{tab:fb_summary} we list values of $t_{\rm fb}$ (Equation~\ref{eq:t_fb}) for all simulations using the gas' properties at $t=t_{\rm end}$, where the semi-major axis of each bound SPH particle, $a = -GM/2E$, is based on its energy, $E=K+U<0$. We use the entire bound mass at the end of the simulations, $M_{\rm fb,end}$ in Table~\ref{tab:fb_summary}, in the knowledge that it is an upper limit, except possibly for the 35M simulation, the only tabulated EoS simulation where the unbound mass starts to plateau in Figure~\ref{fig:evo_vs_time} (lower panel).

An inspection of each SPH particle's $t_{\rm fb}$ (Equation 3) as a function of distance from the centre of mass for our simulations, as done in Paper~I (figure 9) shows that the ideal gas EoS produces a more extended distribution of bound material (the tabulated EoS simulations unbind more surface material reducing the extent of bound gas), up to twice as far as the corresponding tabulated EoS simulation, or $\sim10^4$~\rs, resulting in longer fall-back times (mass-weighted averages of $\sim 200$~yr (ideal gas) vs. $\sim 80$~yr (tabulated); Table~\ref{tab:fb_summary}). A small amount of bound material can remain at large distances for much longer times ($10^4-10^5$~yr).

\begin{figure}
    \centering
    \includegraphics[width=0.98\linewidth]{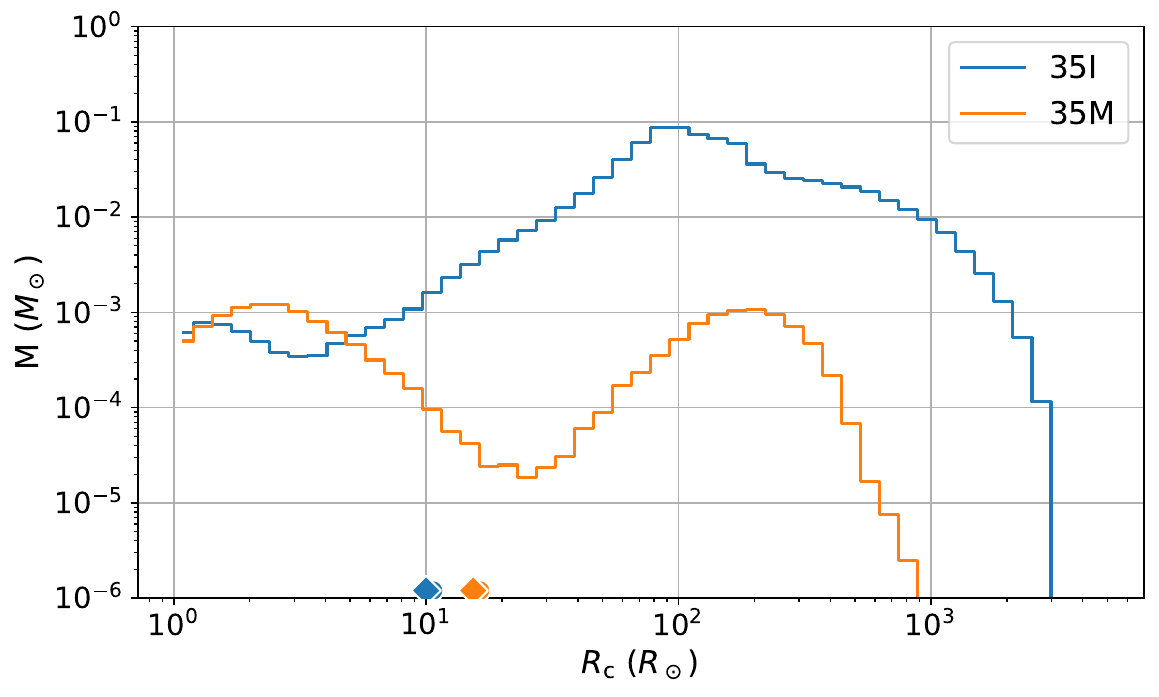}
    \includegraphics[width=0.98\linewidth]{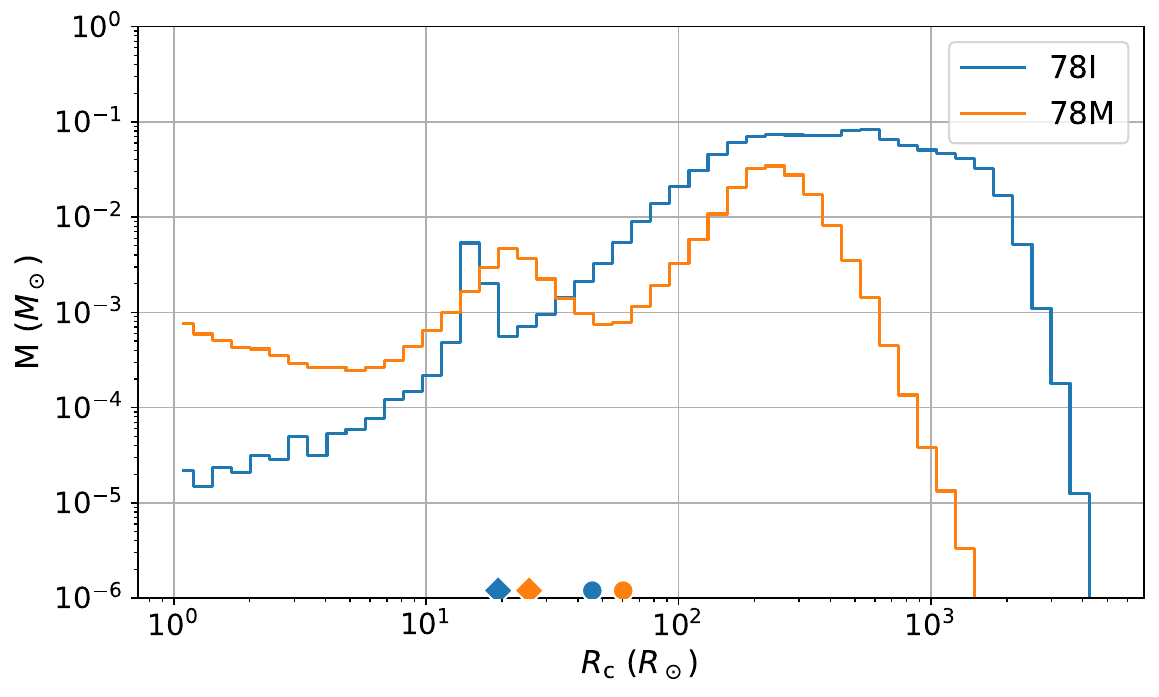}
    \includegraphics[width=0.98\linewidth]{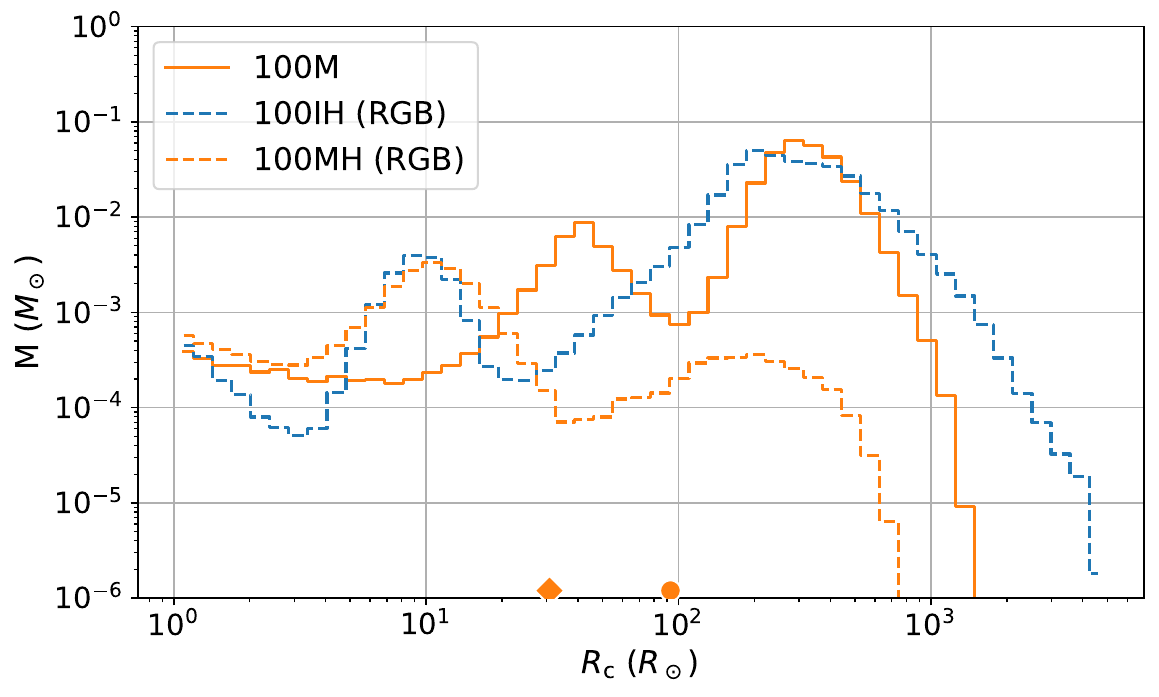}
    \includegraphics[width=0.98\linewidth]{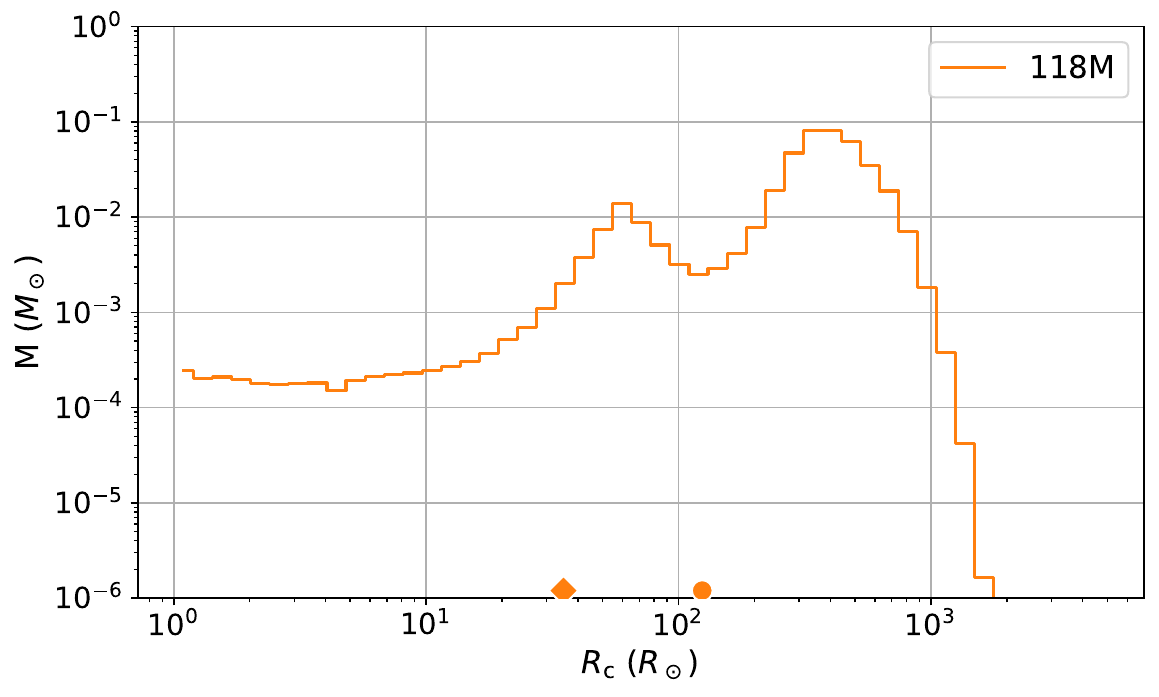}

    \caption{Histograms of bound mass as a function of circularisation radius, $R_c$ (Equation~\ref{eq:Rc}) for, from top to bottom, $q= 0.35$, $q= 0.78$, $q=1.00$, and $q=1.18$. The blue and orange lines are for the ideal gas and tabulated EoS simulations, respectively. Plots are generated at $t=t_{\rm end}$. The circles and diamonds represent the orbital distance from the centre of mass of the system to the primary and companion cores, respectively. Note that for the $q=1.00$ plot we have also included the histograms as found in \citetalias{Nibbs2025} for the 100IH and 100MH simulations (dashed blue and orange lines), respectively.}
    \label{fig:fallback_histogram}
\end{figure}

Figure~\ref{fig:fallback_histogram} shows the bound mass distribution as a function of circularisation radius $R_{\rm c}$, expressing where bound mass from each simulation is expected to fall back to. In all four panels we observe a bimodal distribution: a minor component near $R_{\rm c} \sim d_{\rm orb,1/2}$ (the distance of the two cores from the centre of mass, marked by the circles/diamonds), with too little specific angular momentum to circularise outside the orbit; this material instead remains close to the binary and is more likely to be, eventually, accreted, flung into the stable disk, or even accelerated out of the system on short timescales of 10--100 years \citep{artymowicz_dynamics_1994,lai_circumbinary_2023}. The second, dominant component is a broader peak at $R_{\rm c} \sim 10^{2} - 10^{3}$~\rs\  with sufficient angular momentum  to circularise beyond the binary's orbit.

For the ideal gas simulations, the bound mass in both components exceeds those from the tabulated EoS simulations by roughly two orders of magnitude, consistent with the known lower envelope-unbinding efficiency of ideal gas CE simulations. The bound masses also exceed those reported in \citetalias{Nibbs2025} (Table~\ref{tab:fb_summary}), reflecting the more massive AGB star envelope. The tabulated EoS simulations (orange lines, Figure~\ref{fig:fallback_histogram}) yield significantly lower fallback masses, which are themselves an upper limit, except for the 35M simulation that was evolved long enough to unbind most of the envelope. 

The fall-back disc radii obtained here are larger than those found for the RGB simulations of \citetalias{Nibbs2025} (see also the direct comparison of the $q=1.00$ RGB and AGB simulations in the third panel of Figure~\ref{fig:fallback_histogram}), likely reflecting greater angular momentum payload in the AGB simulations. 

Therefore, at the end of an interaction such as the one we have simulated there, a fall-back and adjustment phase follow, lasting of the order of a couple of centuries during which a cavity develops around the binary and a more stable circumbinary disc is established. Once a dynamically stable disc forms, its long-term evolution is governed by viscous transport on a timescale of $\tau_{\rm visc} = 10^2 - 10^4$~yr. These estimates are derived from Equation~\ref{eq:tVisc}, assuming $\alpha=10^{-3}$ to $10^{-1}$, gas temperatures of $\sim1000~K$ to $10,000~K$ in the region of the binary at the end of the simulations. These values constitute a substantial fraction of the observed circumbinary disc lifetimes of $\sim 10^5$ years \citep{izzard_circumbinary_2023}. 

\begin{table}
    \centering
    \begin{tabular}{ccccc}
    \hline
         Sim.&   $\langle t_{\rm fb}\rangle$& $M_{\rm fb, end}$  & $R_{\rm fb,end}$ &$d_{\rm orb,1/2}$\\
         Name& 
     (yr)& (\ms) & (\rs)&(\rs)\\
     \hline
 35I& 79& $0.76$& 100&11, 10\\
 35M& 194& $0.02$& 210&16, 15\\
 78I& 75& $0.10$& 570&46, 19\\
 78M& 183& $0.12$& 240&67, 28\\
 100M& 201& $0.31$& 290&93, 31\\
 118M& 245& $0.41$& 340&124, 35\\
 \hline
 \end{tabular}
    \caption{Mass-weighted mean fall-back times (Equation~\ref{eq:t_fb}), $\langle t_{\rm fb}\rangle=\sum_i M_{i} t_{{\rm fb},i}/\sum_i M_i$, where $M_i$ is the bound mass contained within the $i$-th fall-back-time bin and $t_{{\rm fb},i}$ is the centre of that bin.  $M_{\rm fb,end}$ is the total mass of the bound envelope at $t=t_{\rm end}$.  $R_{\rm fb,end} \equiv R_c$ of the rightmost peak of each histogram in Figure~\ref{fig:fallback_histogram}. $d_{\rm orb,1/2}$ is the distance of core (1) and companion (2) to the centre of mass  (the circles and diamonds in Figure~\ref{fig:fallback_histogram}, respectively).}
    \label{tab:fb_summary}
\end{table}

\section{Summary}
\label{Sec:Summary}

We have presented a suite of three-dimensional SPH simulations of CE interactions involving an AGB donor across a range of binary mass ratios and two EoS. Our findings are as follows:

\begin{enumerate}
\item Our AGB simulations yield post-interaction separations of $\sim$40–181~\rs (citing the tabulated EoS simulations, with equivalent periods of 27 to 176 days). As such AGB simulations overlap with the smallest periods of the observed post-AGB binaries. These separations are larger than for the RGB counterparts, though the fractional radius reduction is actually similar: for the comparable $q=1.00$ simulations, the final separations (periods) are 141~\rs\ (129~days) and 49~\rs\ (35~days) for the AGB and RGB donors, respectively.  
    
\item All but the two highest mass ratio simulations, and only for ideal gas, go through a CE in the simulated time, while for the RGB simulations of \citetalias{Nibbs2025} only the highest $q=1.5$ simulations (with any equation of state) avoided a conventional (fast) CE inspiral. We are therefore not yet able to state when a CE inspiral can categorically be avoided, although a higher mass ratio clearly is a factor. The timing of the transition between stability and instability appears to be a sensitive function of more than just the mass ratio with a likely strong role of the stellar envelope structure \citep{Webbink1984,Ge2015_II}, which in 3D hydrodynamic simulations remains poorly resolved \citep[e.g.,][]{Mu2026}
   
\item  Simulations that include recombination energy unbind substantially more envelope mass, and result in  wider post-inspiral separations than their ideal gas counterparts. We could not measure accurately the mass that remains bound to make a circumbinary disk. However, the body of evidence from these simulations, those of Paper~I and past experience indicates that the mass available would be between a few hundredths and a tenth of a solar mass. 

\item All our simulations start with circular orbits and acquired a very minor eccentricity. Hence, it is unclear how post-RGB/AGB binaries with substantial eccentricities (e.g., e=0.4-0.6) can result from strong binary interactions. Mechanisms to pump the eccentricity, such as disk-binary interactions \citep[][]{dorazio_orbital_2021,siwek_orbital_2023}, may work for wider systems, but may equally struggle to reproduce observed combination of low orbital periods ($<$200~days) and high eccentricity (0.3-0.4) values, such as some of those observed in Figure~\ref{fig:PvsE_Obs}.  

\item A higher $q$ does not systematically result in a lighter envelope by the time of the in-spiral, as initially hypothesised. Longer pre-inspiral times possible at higher $q$, are usually accompanied by smaller mass loss rates, in a self regulating way. The length of the pre-inspiral phase, during which we measure the mass loss from the primary envelope, is however not a converged property making this statement qualitative. 

\item  Quoting one of our tabulated EoS simulations (with $q=0.35$), if 2\% of the envelope remains bound, $0.023$~\ms\ will fall back to the system. We estimate the fall back times for this bound gas of up to two centuries, pointing to a rapid phase of material return. Due to its angular momentum, most of the fall-back gas would form discs with radii from 0.5 to 5~au. 

\item Our values for $\gamma_{\rm loss}$, the ratio of specific angular momentum lost from the binary, to the specific angular momentum of the binary itself estimated at the onset of inspiral, are $\gamma_{\rm loss}\approx3$--4, for all simulations except the 35I simulation, for which we find $\gamma_{\rm loss} \sim 8$ due to its much smaller mass ratio. There is a very marginal trend of decreasing $\gamma_{\rm loss}$ (more stable mass transfer) with increasing $q$ for the ideal gas simulations, however this is not so clear for the tabulated EoS simulations as the measurement is made uncertain by the short pre-inspiral times in those simulations. This is similar to what is observed for the RGB donors in \citetalias{Nibbs2025}, where that trend is clearer for both ideal gas and tabulated EoS simulations.

\end{enumerate}

\section{Conclusion and Discussion}
\label{sec:conclusion_and_discussion}

With regards to understanding post-giant binary orbital configurations we can infer the following:

{\it Mass transfer stability.} Post-giant systems cannot be the result of regular CE interactions, because their separations are larger than is typical for CE inspirals \citep[][]{Iaconi2019b}. Our RGB (\citetalias{Nibbs2025}) and AGB (this work) interaction simulations attempted to create the conditions for a wider final separation by promoting mass transfer stability before the inspiral, by means of a larger mass ratio. This was predicted to result in a longer phase of pre-inspiral mass transfer and a lighter envelope \citep{Reichardt2019}.  This was not quite born out: higher $q$ simulations {\it do} tend to prolong the pre-inspiral, but not necessarily lighten the envelope more. What does happen, however, is that a higher $q$ could eventually prevent a fast inspiral from occurring altogether, in which case the entire outcome changes -- the diversity of behaviour introduced by the two equations of state makes it difficult to draw firmer conclusions. Ideal gas equation of state (no  recombination energy is released leaving the envelope more compact and promoting stability) or a tabulated equation of state (the entire recombination energy is thermalised such that the envelope is maximally puffed up and promoting mass transfer instability) likely book-end the correct behaviour. Radiation transport is clearly needed to resolve the nature and length of the pre-inspiral phase, and to more accurately represent the outer layers of giant stars in Roche lobe contact.

{\it Post-interaction orbital separation.} Larger mass ratios result in wider CE post-inspiral separations. A  near-unity mass ratio in our AGB binaries leads to separations at the lower end of the observed range of post-giant binaries. It would be useful to study the effect of the envelope mass on the stability of the mass transfer and on the ensuing CE interaction to continue bridging the gap between classic CEs and the observed post-giant binaries. 

{\it Eccentricity.} Our simulations start with a circular orbit and end with  very small eccentricities. The question of how post-giant binaries have eccentricities as high as 0.6 at periods of 1000-2000~days, or as high as 0.3 at periods of 100-700 days is unanswered as any strong binary interaction is expected to circularise the orbit. Alternative mass transfer channel, such as wind RLOF were studied by, e.g., \citet{Krynski2025}, who found that in the presence of a circumbinary disc, binaries undergoing wind RLOF can attain eccentricities of up to $e=0.4$, but only for periods larger than 2000 days. They do not reproduce the most eccentric post-AGB binaries ($e\sim0.4-0.6$), nor do they predict shorter periods systems (eccentric or not). Even the 1D models of \citet{Moltzer2025} that reproduced {\it post-RGB} periods struggle to explain the high eccentricities which in their models need to be retained throughout the period of (stable) mass transfer.

{\it Circumbinary disc.} All post-RGB and AGB binaries have circumbinary discs with typical masses of $\sim 10^{-2}$~\ms\ \citep{kluska_population_2022,oomen_orbital_2018,VanWinckel2025}. Unstable mass transfer before a CE inspiral results in sufficient, bound, $L_2$ mass loss to generate a disc. However, the circumbinary disc is likely swept away by the CE ejecta. In such case, we conclude that a circumbinary disc could form  instead via fall-back of bound material (see also \citealt{kashi_circumbinary_2011,Iaconi2017,Kuruwita2016}) in a few hundred years after the end of the inspiral. Some gas would fall back onto the binary, but some has sufficient angular momentum to come to rest well outside the orbit.  The total mass available to fall back is likely between a few hundreds and about a tenth of a solar mass. Initially, the binary would interact with gas falling back within its orbit, and later it would interact  with a still adjusting circumbinary disc. Only later, as the disc cools and evolves viscously on longer timescales, will it thin out and decrease in mass via accretion. The binary orbital elements may respond to each of these phases \citep[e.g.,][]{Rafikov2016}.

\paragraph{Acknowledgments}
This work was supported by resources awarded under Astronomy Australia Ltd's ASTAC merit allocation scheme on the OzSTAR national facility at Swinburne University of Technology. The OzSTAR program receives funding in part from the Astronomy National Collaborative Research Infrastructure Strategy (NCRIS) allocation provided by the Australian Government, and from the Victorian Higher Education State Investment Fund (VHESIF) provided by the Victorian Government. JN acknowledges funding support from Macquarie University through the Macquarie University Research Excellence Scholarship (MQRES), an Australian Government Research Training Program Scholarship.

\bibliography{references,references_orsola,references_orsola_2}

\appendix 
\renewcommand{\thefigure}{A\arabic{figure}} 
\setcounter{figure}{0}

\section{Conservation Properties}
\label{sec:appendixA}
\begin{figure*}[b]
    \centering
    \includegraphics[width=0.7\linewidth]{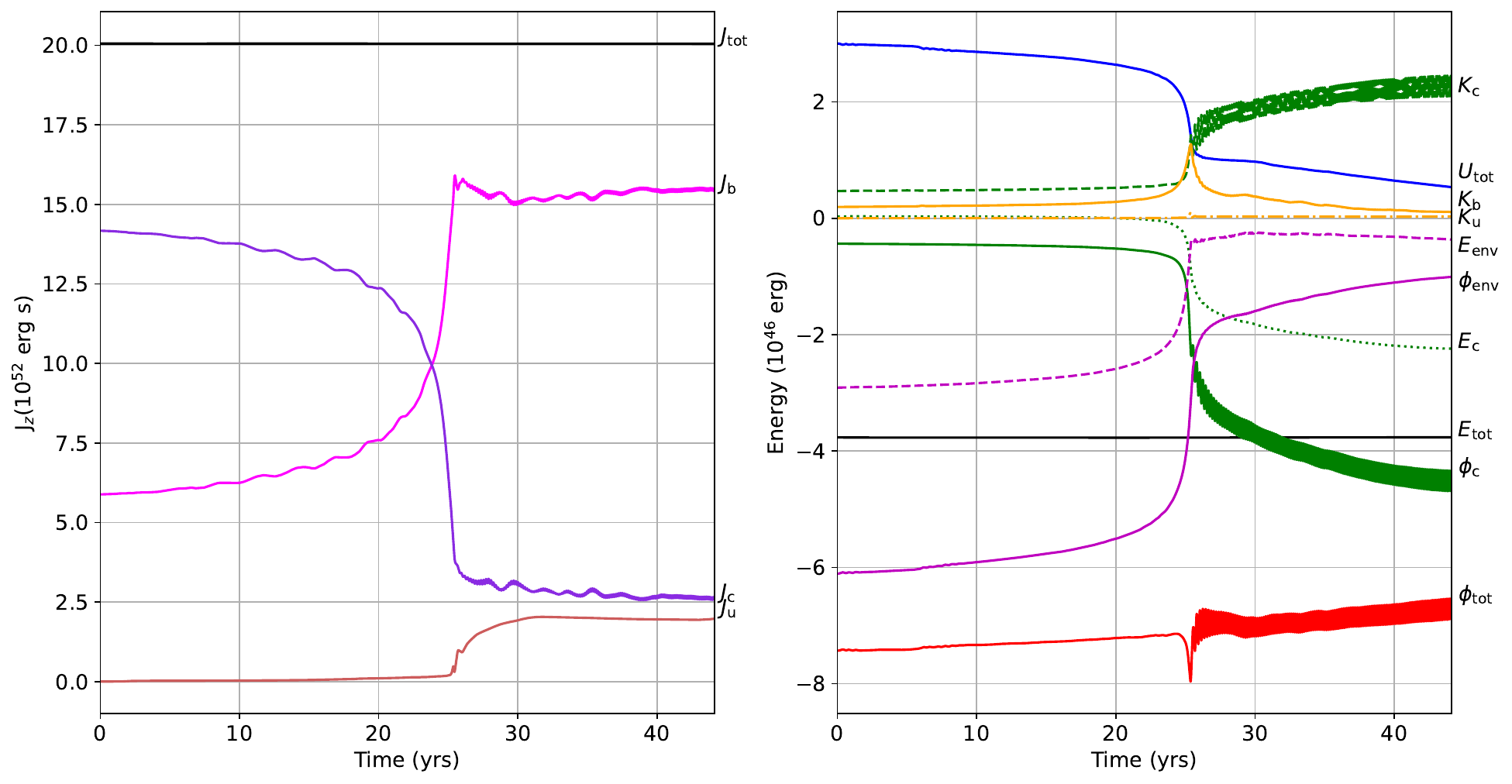}
    \caption{Plots of angular momentum (left) and energy (right) evolution for the 78I simulation, as an example of excellent angular momentum and energy conservation in our simulations. In each plot the subscripts $tot$, $b$, $u$, $c$, $env$, refer to the total, bound, unbound, cores, and gas respectively. In the plot of energy, $K$, $U$, and $\phi$, are the kinetic energy, thermal energy, and the potential energy respectively, while $E$ is used for the combination of potential, kinetic and thermal as appropriate to the cores, envelope or total.}
    \label{fig:conservation}
\end{figure*}

Energy and angular momentum are conserved to excellent precision in all of our simulations. Figure~\ref{fig:conservation} shows the 78I simulation as an example. In all of our simulations momentum is conserved to within 0.2\%, and total energy is conserved to within 0.1\%.

\label{sec:appen1}

\section{The $\gamma$ parameter in Paper~I}
\label{sec:appen3}
\begin{figure*}
    \centering
    \includegraphics[width=0.8\linewidth]{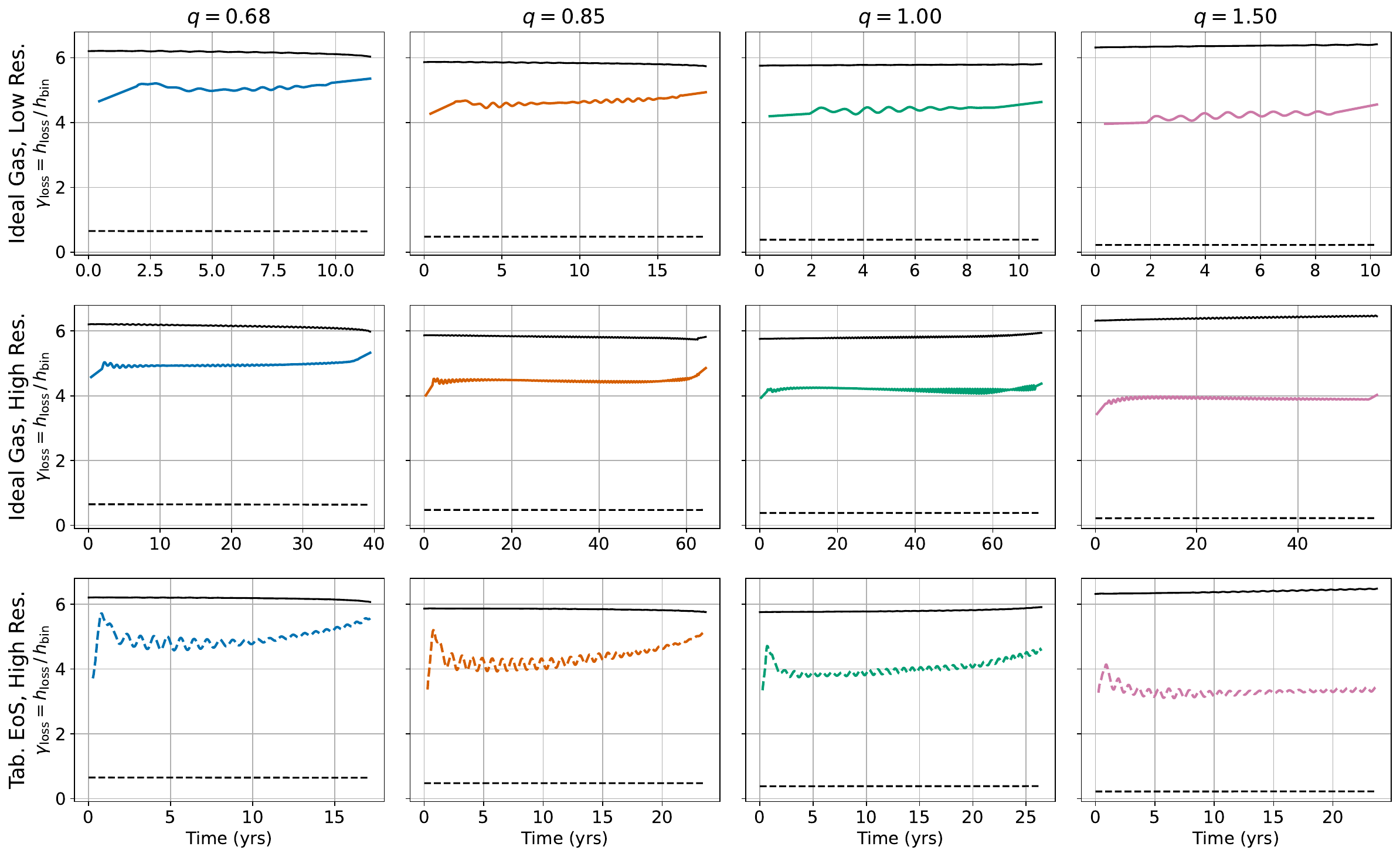}
    \caption{Plots of $\gamma_{\rm loss}$ as a function of pre-inspiral time for each of the simulations in Paper~I. Here the solid black lines denote $\gamma_{\rm L_2}$, whereas the dashed black lines denote $\gamma_{\rm acc}$. In the top row we plot the low resolution ideal gas, 'IL' simulations, followed by the high resolution ideal gas, 'IH' simulationed, and then in the bottom row we plot the tabulated EoS, 'MH' simulations. Each column then shows the corresponding value of mass ratio.}
    \label{fig:gamma_RGB}
\end{figure*}
 
In Paper~I, we have calculated values of $\gamma_{\rm loss}$ and  $\gamma_{\rm L2}$   using only the start of the simulation ($t_0$), and the time of the inspiral's onset ($t_i$), this yielded, e.g., for $\gamma_{\rm loss}$
\begin{equation}
    \gamma_{\rm loss} = \frac{h_{\rm loss}(t=t_{\rm i})}{h_{\rm bin}(t = t_0)}.
\end{equation}
\noindent In this paper 
(Section~\ref{sec:Results}), we instead calculated all values  as a function of time, i.e.
\begin{equation}
    \gamma_{\rm loss} = \frac{h_{\rm loss}(t)}{h_{\rm bin}(t)},
\end{equation}
\noindent and then taken the averages, which leads to a uniformly higher values due to a progressively smaller value  of $h_{\rm bin}$ (as opposed to using only the largest value of $h_{\rm bin}$). 

In Table~\ref{tab:RGB_gamma} we therefore list a revision of the time-average value of all quantities for the RGB simulations of Paper~I, now directly comparable to the values for the AGB stars of this work (Table~\ref{tab:MJ_summary}). The difference for each simulation is at most a 30\% increase and does not affect our conclusions. 
\begin{table}[]
    \centering
    \begin{tabular}{llll}
    \hline
         Sim. Name&   $\langle\gamma_{\rm loss} \rangle$& $\langle\gamma_{L_2} \rangle$&$\langle\gamma_{\rm acc} \rangle$\\
         \hline
         68IL& 5.1& 6.2&0.6\\
 68IH& 5.0& 6.1&0.6\\
 68MH& 5.0& 6.2&0.6\\
 85IL& 4.6& 5.8&0.5\\
 85IH& 4.5& 5.8&0.5\\
 85MH& 4.4& 5.8&0.5\\
 100IL& 4.4& 5.8&0.4\\
 100IH& 4.2& 5.8&0.4\\
 100MH& 4.0& 5.8&0.4\\
 150IL& 4.4& 6.6&0.2\\
 150IH& 3.9& 6.4&0.2\\
 150MH& 3.3& 6.4&0.2\\
 \hline
 \end{tabular}
    \caption{Revision of table~4 from Paper~I using the values of $\langle\gamma\rangle$ calculated taking both $h_{\rm bin}$ and $h_{\rm loss}$ as functions of time. }
    \label{tab:RGB_gamma}
\end{table}





\end{document}